\documentclass[prl,reprint,superscriptaddress,nofootinbib]{revtex4-1}

\usepackage[linktoc=page,colorlinks,urlcolor=blue,citecolor=blue,linkcolor=blue]{hyperref}
\usepackage{amssymb,amsmath}
\usepackage{graphicx}
\usepackage{natbib}
\usepackage{mathrsfs}

\usepackage{ulem}

\usepackage{color}
\usepackage{upgreek}
\usepackage[letterpaper,textwidth=7in,top=.75in,bottom=.75in]{geometry}

\newcommand{\unm}{Center for High Technology Materials and Department of Physics and Astronomy, University of New Mexico, Albuquerque, NM, USA}
\newcommand{\fom}{Fomblin\textsuperscript{\textregistered}}
\begin{document}
\title{Solution nuclear magnetic resonance spectroscopy on a nanostructured diamond chip}
\date{\today}
\author{P. Kehayias$^{\mathsection}$}
\affiliation{Department of Physics, Harvard University, Cambridge, MA, USA}
\affiliation{\unm}

\author{A. Jarmola$^{\mathsection}$}
\email{andrey.jarmola@odmrtechnologies.com}
\affiliation{ODMR Technologies Inc., El Cerrito, CA, USA}
\affiliation{Department of Physics, University of California-Berkeley, Berkeley, CA, USA}

\author{N. Mosavian}
\affiliation{\unm}

\author{I. Fescenko}
\affiliation{\unm}

\author{F. M. Benito}
\affiliation{\unm}

\author{A. Laraoui}
\affiliation{\unm}

\author{J. Smits}
\affiliation{\unm}

\author{L. Bougas}
\affiliation{Johannes Gutenberg Universit\"{a}t Mainz, 55128 Mainz, Germany}

\author{D. Budker}
\affiliation{ODMR Technologies Inc., El Cerrito, CA, USA}
\affiliation{Department of Physics, University of California-Berkeley, Berkeley, CA, USA}
\affiliation{Helmholtz Institut Mainz, 55099 Mainz, Germany}

\author{A. Neumann}
\affiliation{\unm}

\author{S. R. J. Brueck}
\affiliation{\unm}

\author{V. M. Acosta}
\email{vmacosta@unm.edu}
\affiliation{\unm}

\renewcommand{\thefootnote}{}{\footnote{$\mathsection$ Equal contribution}}

\begin{abstract}
We demonstrate nuclear magnetic resonance (NMR) spectroscopy of picoliter-volume solutions with a nanostructured diamond chip. Using optical interferometric lithography, diamond surfaces were nanostructured with dense, high-aspect-ratio nanogratings, enhancing the surface area by more than a factor of 15 over mm$^2$ regions of the chip. The nanograting sidewalls were doped with nitrogen-vacancy (NV) centers so that more than 10 million NV centers in a (25 $\upmu$m)$^2$ laser spot are located close enough to the diamond surface ($\sim5$ nm) to detect the NMR spectrum of $\sim1$ pL of fluid lying within adjacent nanograting grooves. The platform was used to perform  $^{1}$H and $^{19}$F NMR spectroscopy at room temperature in magnetic fields below 50 mT. Using a solution of CsF in glycerol, we demonstrate that $4\pm2\times10^{12}$ $^{19}$F spins in a $\sim1~{\rm pL}$ volume, can be detected with a signal-to-noise ratio of 3 in 1 s integration. This represents nearly two orders of magnitude improvement in concentration sensitivity over previous NV and picoliter NMR studies.
\end{abstract}
\maketitle
Nuclear magnetic resonance (NMR) spectroscopy is an invaluable analytical tool  for determining the composition, structure, and function of complex molecules. However, conventional high-field NMR spectrometers have drawbacks. Due to their reliance on cryogenic magnets, the best NMR spectrometers are massive, immobile, and expensive. Often they require relatively large quantities of analyte and large ($\gtrsim$10 $\upmu$L) sample volumes to overcome fundamental sensitivity constraints from inductive detection and low thermal polarization. This limits  NMR use in sample-limited analysis and high-throughput screening, where a parallel microfluidic platform would be preferable \cite{Lee2008, Ha2014, Rubakhin2011}. Microcoil NMR \cite{fratilaReview, laceyReview, Olson1995, suterMicroslot, mcDowellMicrocoil} can lower the required analyte volume to $\gtrsim1~$nL, but still requires a large magnetic field for sensitive spectroscopy. Nuclear hyperpolarization methods and alternative magnetometry technologies \cite{micahZULFreview, clarkeSQUIDmriReview}  enable certain experiments at lower fields, but there is still no platform that combines the high sensitivity and sub-nL analyte volume necessary for operation in microfluidic assays. 

Recently, a new technique has emerged for NMR spectroscopy at the nanometer scale based on optical detection of the electron spin resonances of negatively-charged nitrogen-vacancy (NV) color centers in diamond \cite{wrachtrupPMMA, Mamin_PMMA, devience_nuclSens, rugar2015, Haeberle2015, spuriousHarmonics, merilesDiffusion, igorProteinNMR}. The technique relies on statistical nuclear polarization, which is substantially larger than thermal polarization for nanoscale sensing volumes \cite{herzog_thermVsStat}. Combined with non-inductive optical detection, this renders the NV NMR sensitivity independent of temperature and magnetic field. Previous works used single NV centers \cite{wrachtrupPMMA, Mamin_PMMA, devience_nuclSens, rugar2015, Haeberle2015, spuriousHarmonics, merilesDiffusion}, or an ensemble \cite{devience_nuclSens}, to detect NMR of nuclei in liquids and thin films across a flat sample-sensor interface. While this was a significant scientific breakthrough, the long measurement time (hours to days) are a liability for many applications.
\begin{figure}[ht]
\begin{center}
\includegraphics[width=0.48\textwidth]{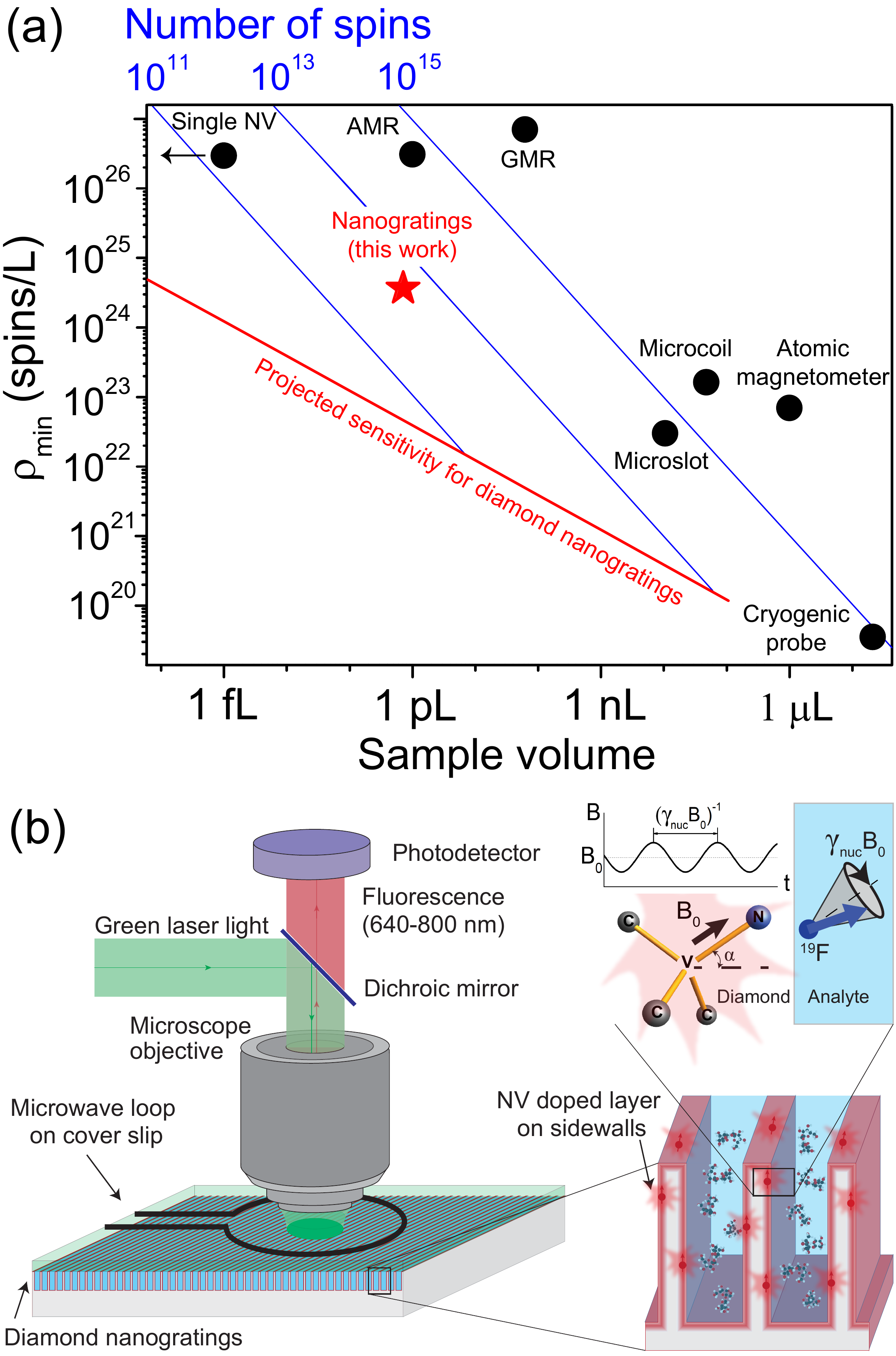}
\end{center}
\caption{\label{fig1} Picoliter NMR. (a) Overview of ambient-temperature NMR techniques for small volumes. Points represent experimental values for minimum detectable nuclear-spin concentration in 1 s with SNR = 3 for microslot \cite{suterMicroslot}, microcoil \cite{mcDowellMicrocoil}, and cryogenic \cite{kovacsCryoProbeReview} probes, atomic vapor magnetometers \cite{micah2008}, giant magnetoresistance sensors (GMR) \cite{gmr_nmr}, anistropic magnetoresistance sensors (AMR) \cite{amr_nmr}, single NV centers \cite{rugar2015}, and NV-doped nanogratings (this work).  The solid red line is the projected sensitivity for diamond nanogratings [Eq.~\eqref{eq:1}], exhibiting volume$^{-1/2}$ scaling \cite{suppl}. Solid blue lines indicate constant numbers of spins. (b) Epifluorescence diamond NMR setup. First inset: the sensor region consists of dense, high aspect-ratio diamond nanogratings fabricated via interferometric lithography and doped with NV centers. Second inset: experimental geometry. The analyte's precessing nuclear statistical polarization produces an oscillating magnetic field which is sensed by adjacent near-surface NV centers. 
}
\end{figure}

In this work we demonstrate picoliter (pL) solution NMR  using a nanostructured diamond chip. The NMR detection sensitivity depends on the number of NV centers that are located close enough to the diamond surface to sense external spins. To increase this number, the diamond surface was lithographically structured with dense, high-aspect-ratio nanogratings to enhance the sensor-analyte contact area by $\gtrsim15\times$. The nanostructure sidewalls were then doped with a high density of NV centers. The result is tens of millions of NV centers are located close enough to the diamond surface (5-20 nm) to detect the NMR spectrum from $\sim1~$pL of fluid lying within the adjacent nanograting grooves. This leads to a corresponding boost in fluorescence signal and reduction in NMR acquisition time. With further improvements in spectral resolution, this platform could enable a wide variety of applications in biochemistry, including pharmacodynamic studies of metabolites and natural products and high-throughput screening for drug discovery.

NMR is a powerful analytical technique for non-destructive molecular structure elucidation, but its detection sensitivity is orders of magnitude worse than other analytical chemistry techniques such as mass spectrometry or fluorescence labeling methods \cite{laceyReview}. The sensitivity is limited by the small nuclear magnetization. At the highest DC magnetic field available, $B_0=24~$T, the room-temperature $^1$H thermal polarization is just $10^{-4}$ \cite{24Tmagnet}. The sensitivity is further limited by frequency-dependent noise due to inductive detection. One potential remedy is to increase the magnetic field, but despite steady improvements in magnet technology, the signal strength has improved by less than a factor of two over the last 20 years \cite{Polenova2016}. 

Alternative NMR techniques seek to increase nuclear polarization and/or lower readout noise without relying on increasing $B_0$. The latter can be accomplished by cryogenically cooling the inductive probe \cite{kovacsCryoProbeReview} or switching to \textit{non-inductive detection} modalities, including GMR \cite{gmr_nmr}, AMR \cite{amr_nmr}, or atomic magnetometers \cite{micah2008}. The present NV-based NMR approach uses non-inductive detection to sense the \textit{statistical nuclear polarization}, an effect which arises from imperfect cancellation of the net magnetization from an ensemble of randomly oriented spins \cite{wrachtrupPMMA, Mamin_PMMA}. Statistical polarization is larger than thermal polarization for sufficiently small numbers of nuclear spins \cite{suppl} and makes the NV NMR sensitivity independent of sample temperature and $B_0$. Figure \ref{fig1}(a) summarizes existing NMR techniques for small sample volumes. The NMR sensitivity is characterized by the minimum spin concentration detectable in 1 s at room temperature with signal-to-noise ratio (SNR) of 3. 

For NV NMR the minimum detectable spin concentration $\rho_{min}$ for SNR$=3$ in 1 s is  \cite{suppl}:
\begin{equation} \label{eq:1}
\rho_{min} = \frac{3}{P(\alpha)(\mu_0 \hbar \gamma_{NV}\gamma_{nucl})^2}\times \frac{d_{NV}^3}{T_{tot}^2C\sqrt{\eta N_{NV} N_r}},
\end{equation}
where $P(\alpha)=\{\pi[8-3(\sin\alpha)^4]\}/128$ is a geometric factor that comes from the angle $\alpha$ the N-V axis makes with the diamond surface normal [Fig.\ref{fig1}(b)], $\mu_0=4\pi\times10^{-7}$ m$\cdot$T/A is the vacuum permeability, $\hbar=1.055 \times 10^{-34}$ J$\cdot$s is the reduced Planck constant,  $\gamma_{NV}=28.03~$GHz/T is the NV gyromagnetic ratio, $\gamma_{nucl}$ is the nuclear gyromagnetic ratio (42.58 MHz/T for $^1$H and $40.08$ MHz/T for $^{19}$F), $d_{NV}$ is the characteristic NV depth below the diamond surface, $T_{tot}$ is the NV phase accumulation time during a single XY8-N pulse sequence [Fig. \ref{corrspecResults}(a)], $C$ is the NV fluorescence-detected spin contrast, $N_{NV}$ is the number of near-surface NV centers in the sensing area, $\eta$ is the mean number of photons collected per NV per readout ($\eta<1$), and $N_r$ is the number of readouts per second. 

Since $\rho_{min}\propto1/\sqrt{\eta N_{NV}}$, the sensitivity can be improved by increasing the sensor surface area, boosting $N_{NV}$ for a constant laser spot size. To this end, we nanostructured the diamond surface with high-aspect-ratio nanogratings and doped the sidewalls with a high density of NV centers.

\begin{figure}[ht]
\begin{center}
\includegraphics[width=0.48\textwidth]{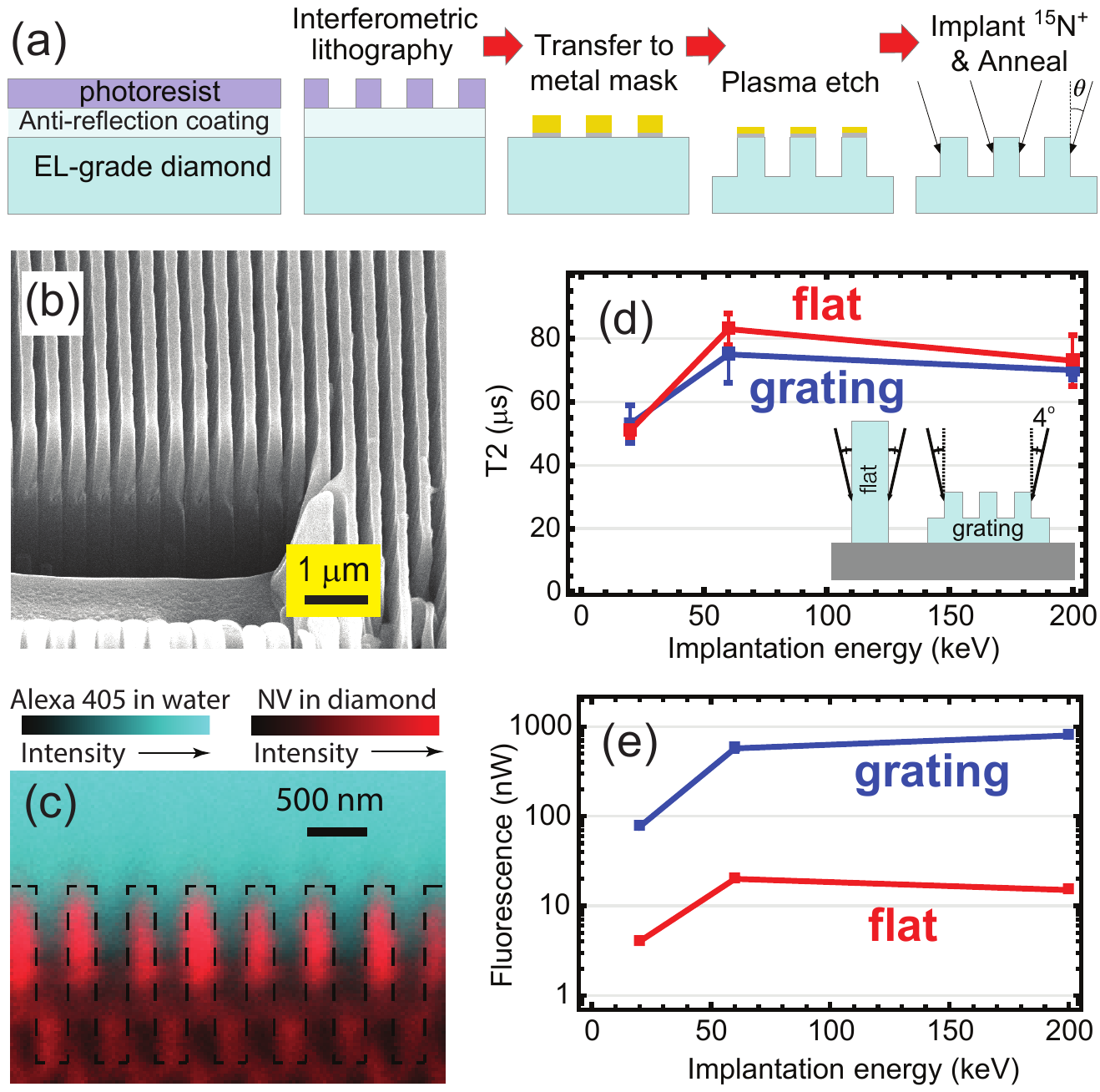}
\end{center}
\caption{ \label{fig2} Diamond nanogratings. (a) Schematic of large-area nanofabrication process. (b) Scanning electron micrograph of 400-nm pitch diamond nanogratings. Focused ion beam etching prior to imaging enabled visualization of the nanogratings cross section. (c) Confocal microscopy images reveal that fluorescence from dye-stained water originates from areas inside the nanograting grooves, confirming wetting. Dashed lines represent the estimated diamond-water boundary. (d) Comparison of $T_2$, measured with the XY8-22 protocol, and (e) fluorescence intensity between flat and nanograting chips implanted at similar conditions.
}
\end{figure}

\begin{figure}[ht]
\begin{center}
\includegraphics[width=0.48\textwidth]{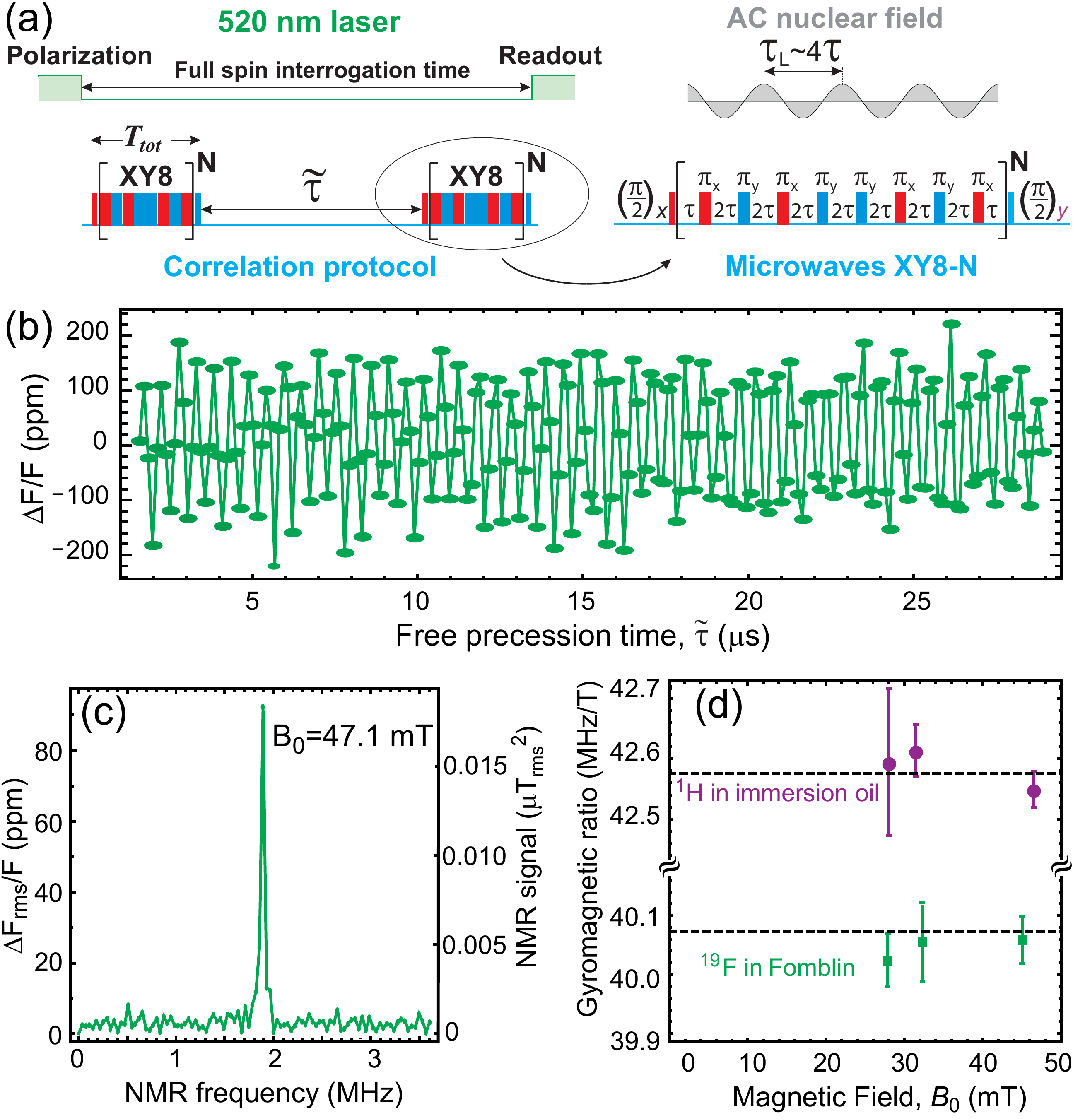}
\end{center}
\caption{ \label{corrspecResults} Nanoscale NMR. (a) Sensing protocols: optical pulses are used to pump and probe NV spin state via the spin-dependent fluorescence; microwave multipulse sequences are applied between optical pump and probe pulses. Red and blue color indicates different microwave phases which are shifted relative to each other by 90$^{\circ}$. NV centers are resonantly tuned to detect a particular nuclear species by setting $4\tau=\tau_L$, where 2$\tau$ is the separation between $\pi$ pulses and $\tau_L$ is the nuclear precession period. In order to reject common-mode noise the sequences are repeated with the phase of the last $\pi$/2 pulse shifted by 180$^{\circ}$. The resulting signals are then subtracted and normalized to give the measurement results. (b) Time-domain NMR signal for $^{19}$F nuclei in \fom~oil taken using XY8-13 correlation sequence. (c) $^{19}$F frequency-domain NMR signal obtained by Fourier Transform of the data in (b).  (d) Measured $^1$H and $^{19}$F gyromagnetic ratios at different $B_0$ values. Dashed lines are literature values \cite{stone2014}.
}
\end{figure}

Nanogratings were etched into the surface of electronic-grade [100]-polished diamond chips using optical interferometric lithography \cite{Brueck2005, Xia2011} and diamond plasma etching \cite{Faraon2013, Hausmann2010, suppl}, as shown in Fig.~\ref{fig2}(a). The nanogratings have a 400 nm pitch and a depth up to $3~\upmu$m [Fig.~\ref{fig2}(b)]. By varying the resist, postbake, and development conditions, duty cycles of $20\mbox{-}80\%$ are achievable. The nanograting sidewalls were doped with NV centers by implanting $^{15}$N$^+$ ions at angles $\theta =\pm 4^{\circ}$ relative to the substrate surface normal. The angles were chosen to ensure an entire $3\mbox{-}\upmu$m-tall sidewall would be doped. Each chip was implanted with either 20, 60, or 200 keV ion energy, corresponding to approximately 5, 10, and 20 nm simulated NV depths, respectively \cite{suppl}. Doses and other information are listed in Tab. S1 in \cite{suppl}. After implantation, the chips were annealed in vacuum at 800-1100 $^{\circ}$C to form NV centers \cite{Acosta2009,suppl}. We investigated the analyte/nanograting adhesion (wetting) with confocal microscopy \cite{suppl}. Water stained with Alexa 405 dye was dispersed on top of an NV-doped nanograting chip. Afterwards, fluorescence from the NV centers (650-800 nm) and dye-stained water (425-500 nm) was simultaneously imaged. We confirmed that the nanogratings were wetting by observing Alexa 405 fluorescence from areas inside the nanograting grooves, Fig.~\ref{fig2}(c).

Optical NMR detection was performed with a custom-built epifluorescence microscope with pulsed laser and microwave interrogation, Fig.~\ref{fig1}(b) \cite{suppl}. Using $\sim140~$mW of 520 nm laser light over a ($\sim25~\upmu$m)$^2$ spot, fluorescence from $\gtrsim10$ million NV centers adjacent to $\sim1~$pL of analyte was detected. A static magnetic field $B_0 =$ 20-50 mT was applied along one of the four NV axes with a permanent magnet, and the aligned NV sub-ensemble was optically interrogated using resonant microwaves ($10\mbox{-}15$ MHz Rabi frequency) delivered by a copper loop fabricated on a cover slip.

To assess the anticipated improvement of the nanograting chips, the nanograting sidewalls and the flat surfaces of unstructured diamond chips were doped at similar conditions to have the same nitrogen density, Fig.~\ref{fig2}(d) inset. Figures \ref{fig2}(d) and (e) compare the NV transverse spin coherence time $T_2$ (measured with the XY8-22 pulse sequence \cite{suppl, deLange2010}) and fluorescence intensity for flat and nanograting chips implanted at different energies. While $T_2$ for flat and nanograting chips is approximately the same (and is similar to $T_2$ for normal-incidence high-dose implantation \cite{devience_nuclSens}), the fluorescence intensity is $20\mbox{-}50$ times brighter for nanograting chips. We attribute this to the $2/\tan4^{\circ}=28$ times higher effective dose captured by the nanograting chip and improved collection efficiency from nanostructures \cite{Rondin2014}. These results highlight the advantage of nanogratings in NMR sensitivity, since fluorescence intensity is proportional to $\eta N_{NV}$ whereas $\rho_{min}\propto1/\sqrt{\eta N_{NV}}$. The nanostructuring process does not negatively affect other parameters in Eq.~\eqref{eq:1}. The fluorescence contrast of Rabi oscillations was $2.7\pm0.8\%$ for all studied diamond chips independent of nanofabrication or doping parameters \cite{suppl}.

%\section{NV NMR spectroscopy}
We used a correlation-spectroscopy pulse sequence for NV NMR detection of external solutions \cite{c13CorrSpec}, as described in Fig.~\ref{corrspecResults}(a). This sequence correlates the nuclear magnetic fields at different points in time, encoding this information in the NV spin state and corresponding fluorescence intensity. Specifically, we used two XY8-N pulse trains, both tuned to the target nuclear Larmor frequency and separated by a variable delay $\tilde{\tau}$. As $\tilde{\tau}$ is swept, the relative NV fluorescence intensity, $\Delta F/F$, oscillates at the nuclear Larmor frequency, analogous to a nuclear free induction decay (FID). The Fourier transform of this FID-like signal reveals the NMR spectrum, from which we extract the spin density. 

Figures~\ref{corrspecResults}(b,c) show the time- and frequency-domain NMR signals for $^{19}$F nuclei in \fom~oil (6600 Da) using the 20 keV nanograting chip. The sensor response was converted to absolute units of nT$^2$ using analytical expressions \cite{suppl} that were validated using calibrated magnetic fields from a test coil. To confirm that the signals arise from the target nuclei, we repeated these measurements at different magnetic fields and with $^{1}$H-rich analytes (glycerol and immersion oil). The resulting NMR peaks were always at the expected Larmor frequency of each species [Fig.~\ref{corrspecResults}(d)] and NMR peaks were either absent (for $^{19}$F) or greatly diminished (for $^{1}$H \cite{devience_nuclSens, rugar2015}) when the analyte was removed.
\begin{figure}[ht]
\begin{center}
\includegraphics[width=0.48\textwidth]{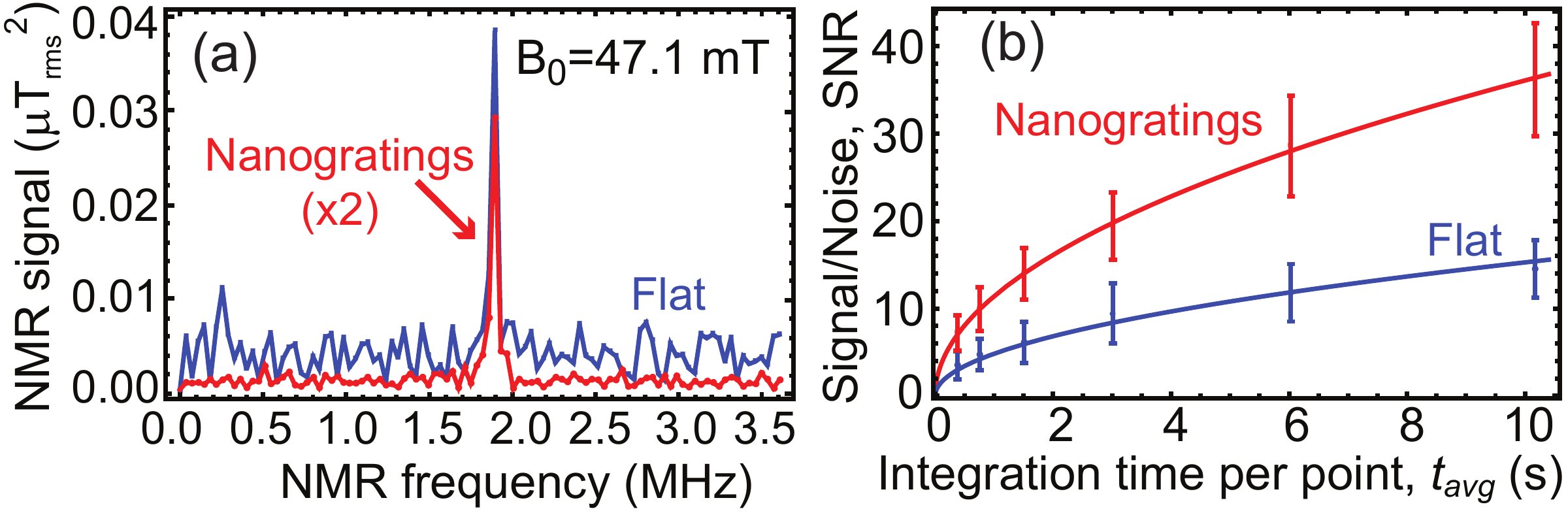}
\end{center}
\caption{ \label{fig4} NV NMR sensitivity characterization. (a) $^{19}$F NMR signal from \fom~oil for flat and nanograting sensors implanted at 20 keV using XY8-13 correlation sequence. (b) NMR signal-to-noise ratio as a function of averaging time. Error bars represent the standard deviation of multiple data sets. Fits to the function SNR=$\alpha\sqrt{t_{avg}}$ give good agreement for both sensors (solid lines). The coefficient $\alpha$ was 2.4 times larger for the nanograting chip.
}
\end{figure}

To characterize the spin-concentration sensitivity, we continuously acquired NMR spectra of $^{19}$F nuclei in \fom~oil and compared SNR for both flat and nanograting sensors implanted at 20 keV. Figure \ref{fig4}(a) shows characteristic NMR spectra for each. The nanograting-sensor noise, defined as the standard deviation of points adjacent to the NMR peak, is a factor of 6 smaller than the flat sensor noise. This is primarily due to the approximately $20\times$ larger fluorescence intensity [Fig.~\ref{fig2}(e)] exhibited by the nanograting sensor, which leads to smaller relative photon shot noise. However, the nanograting sensor signal strength, defined as the $^{19}$F NMR peak amplitude in nT$^2$, is approximately $2.5\times$ smaller. This was unexpected;  the signal amplitude should only depend  on the NV depth, which should be the same for both sensors under identical implantation conditions (Fig.~\ref{fig2}(d) inset). A likely cause for this discrepancy is that the tops of the nanogratings were inadvertently implanted due to degradation of the etch mask [Fig.~\ref{fig2}(a)]. Ions bombarding the nanograting tops are nearly normally incident to the surface, resulting in NV centers that are too deep to sense external nuclei, reducing the overall NMR signal contrast \cite{suppl}. If $50\%$ of the NV centers were formed from deep implantation into the nanograting tops, we would expect a two-fold reduction in signal. Other contributions may be from deep implantation into the flat bottom of the nanogratings or imperfect wetting. 

Regardless, the nanograting sensors had better overall SNR. To acquire each spectrum, $\tilde{\tau}$ was swept and the signal was averaged for a variable time $t_{avg}$. Figure \ref{fig4}(b) plots the SNR as a function of $t_{avg}$, revealing a $2.4\times$ SNR improvement with the nanogratings. \fom~has $40\pm2\times10^{24}$ $^{19}$F spins/L, and we detect it with SNR$=11.4\pm0.2$ at $t_{avg}=1~$s. We thus determine a minimum detectable concentration $\rho_{min}=11\pm1\times10^{24}$ spins/L. Throughout we use $t_{avg}$ as the effective integration time, since the number of points used to generate the spectrum may be reduced using optimized sampling strategies \cite{Scheuer2015}.

\begin{figure}[ht]
\begin{center}
\includegraphics[width=0.4\textwidth]{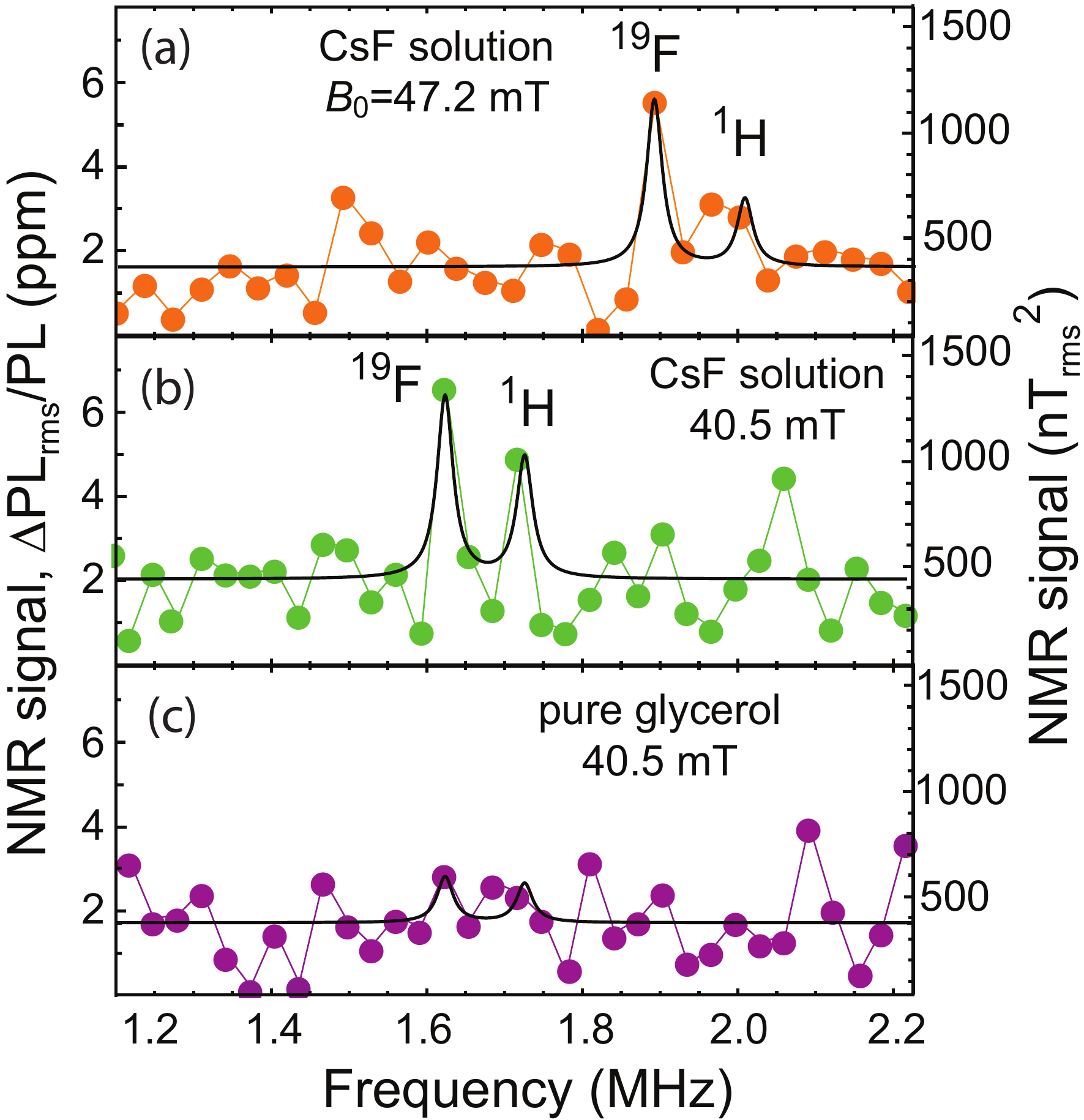}
\end{center}
\caption{ \label{fig5} Solution NMR. (a,b) NMR spectra of $20\%$ by weight CsF solution in glycerol at 47.2 mT and 40.5 mT (c) NMR spectrum of pure glycerol. All spectra were measured with an XY8-10 correlation spectroscopy pulse sequence tuned to the $^{19}$F precession frequency. Solid black lines are fits to Lorentzian peaks with the linewidth and central position fixed based on the known correlation time and gyromagnetic ratios, respectively; amplitudes and offset were the only variable parameters.}
\end{figure}
Next, we performed NMR spectroscopy on more dilute solutions to demonstrate our new capability. We selected CsF dissolved in glycerol as our test analyte. Glycerol was chosen as the solvent because of its high viscosity, which limits molecular diffusion, while CsF was selected as the target molecule due to its high solubility in glycerol and because $^{19}$F has the second largest magnetic moment amongst common nuclear isotopes. Protons have the largest magnetic moment but were not suitable because a background proton signal was often present \cite{devience_nuclSens, rugar2015}. Using a $20\%$ CsF/glycerol solution by weight, the $^{19}$F concentration is $1.0\times10^{24}$ spins/L, about 40 times lower than \fom. To successfully obtain a spectrum with the same SNR requires $\sim$1600 times more signal averaging, a task previously not practical in NV NMR. 

Figure \ref{fig5} shows NMR spectra for this solution. At $B_0=47.2~$mT, we observe a peak at the $^{19}$F Larmor frequency. When changing the magnetic field to $B_0=40.5~$mT, the peak moves according to the $^{19}$F gyromagnetic ratio and maintains a comparable amplitude. Finally, when the analyte is replaced with pure glycerol the peak disappears, as expected. In some cases, hints of a smaller peak at the $^{1}$H Larmor frequency are observed with an amplitude that is consistent with the tails of the XY8 sequence's filter function \cite{c13CorrSpec,spuriousHarmonics}. For the $^{19}$F peak, the SNR is $4\pm1$ at $t_{avg}=77~$s, which corresponds to a minimum detectable concentration $\rho_{min}=6\pm2\times10^{24}$ spins/L. This is roughly a factor of 2 better than for \fom measurements~due to more optimal readout timing.

\begin{figure}[ht]
\begin{center}
\includegraphics[width=0.48\textwidth]{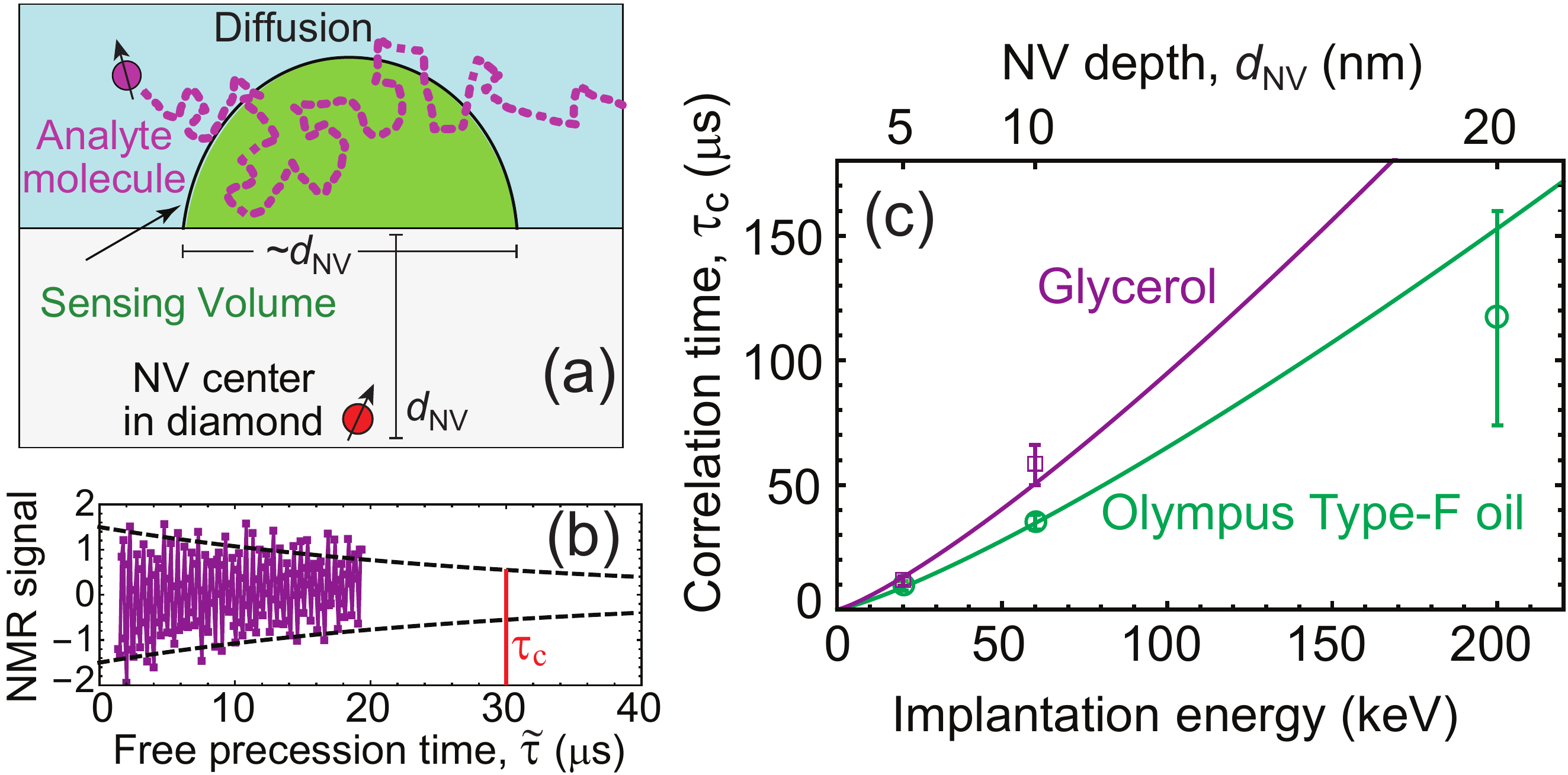}
\end{center}
\caption{ \label{fig6} Diffusion-limited NMR. (a) Diffusion of analyte molecules through the sensing volume. (b) Example of temporal decay of the correlation signal for immersion oil. The decay is exponential with a characteristic correlation time $\tau_C$. (c) Correlation time $\tau_C$ as a function of nitrogen implantation energy for immersion oil and glycerol. Error bars represent fit uncertainty. Solid lines are fits to the one-sided diffusion model discussed in the text.}
\end{figure}

The present platform still has significant room for improvement in sensitivity and spectral resolution. The observed sensitivity is about two orders of magnitude worse than the theoretical sensitivity calculated with Eq.~\eqref{eq:1} and plotted in Fig.~\ref{fig1}(a). The largest contributions to this discrepancy are \cite{suppl}: NV control pulse errors \cite{nirDDarb}, which reduce the contrast by $\sim5\times$ from the ideal case; imperfect doping parameters, which reduce $N_{NV}$ by up to $5\times$ and increase $d_{NV}$ by up to $2\times$ from the ideal case; and the previously-discussed $\sim2.5\times$ lower signal strength for nanograting samples. The spectral resolution, which is inversely proportional to the analyte correlation time $\tau_C$, is currently limited to a few kHz. For nanoscale sensing volumes, analyte molecules rapidly diffuse across the sensing region [Fig.~\ref{fig6}(a)], limiting $\tau_C$. Figure \ref{fig6}(b) plots a typical correlation signal that decays exponentially with time constant, $\tau_C$, in the microsecond range. Figure \ref{fig6}(c) plots $\tau_C$ of protons in immersion oil and glycerol as a function of NV layer depth $d_{NV}$ \cite{suppl}. The data are fit with a one-sided diffusion model, $\tau_C= 2 d_{NV}^2 / D$ \cite{linhNMRdepth}, where $D$ is the molecular diffusion coefficient, revealing $D_{\rm glyc} = 4.4\pm0.6\times10^{-12}$ m$^2$/s and $D_{\rm oil} = 6.4\pm0.3\times10^{-12}$ m$^2$/s. These values are slightly higher than room-temperature literature values ($D_{\rm glyc} =2.5 \times10^{-12}$ m$^2$/s \cite{Tomlinson1973} and $D_{\rm oil} = 0.5\mbox{-}2.5\times10^{-12}$ m$^2$/s) \cite{linhNMRdepth,loretzAPL}, which may be attributed to elevated analyte temperature or hygroscopic effects. The qualitative agreement supports the hypothesis that diffusion is responsible for the short correlation times and improves confidence in the NV depths reported by simulations \cite{suppl}. Lowering the temperature or using microporous media \cite{Li1999} would restrict translational diffusion, improving the resolution and $\rho_{min}$, though dipolar broadening may be a limitation. Alternatively the use of dynamic nuclear polarization may enable NV NMR with nuclear-$T_1$-limited resolution and improved sensitivity due to coherent nuclear precession \cite{Abrams2014, London2013}.

Nevertheless, the current platform's record sensitivity makes it promising for solid-state NMR and nuclear quadrupole resonance (NQR) spectroscopy of trace powders and thin films \cite{Yesinowski1995, Garroway2001}, since these applications require only coarse frequency resolution and diffusion is restricted. It may also find application in NMR relaxometry \cite{Demas2011} and solution NMR for impurity profiling and quality control of pharmaceuticals \cite{Maggio2014}. These applications could benefit from the ease of microfluidic integration, which would permit parallel measurements. Future implementations may also harness the optical waveguding properties of the nanogratings to reduce excitation intensity and increase fluorescence collection by exciting and collecting light through the sides of the chip.

In summary, we performed NMR spectroscopy of picoliter-volume solutions using a nanostructured diamond-chip platform. The sensor uses non-inductive detection of statistical polarization, avoiding the need for large magnetic fields or hyper-polarization. Etching dense, high-aspect-ratio nanogratings into the diamond surface resulted in a 15-fold improvement in surface area and more than 20-fold increase in fluorescence intensity without sacrificing the NV spin properties. Using a solution of CsF in glycerol, we determined $4\pm2\times10^{12}$ spins in a $\sim1~$pL volume, can be detected with a signal-to-noise ratio of 3 in 1 s integration. This represents nearly two orders of magnitude improvement over previous picoliter NMR demonstrations.

\section{Acknowledgements}
We gratefully acknowledge advice and support from C. Santori, Z. Huang, D. Twitchen, D. Suter, D. Bucher, D. Glenn, R. Walsworth, F. Hubert, and Y. Silani. This work was made possible by funding from NSF (IIP \#1549836) and NIH (1R41GM119925-01). P.K. acknowledges support from the Intelligence Community Postdoctoral Research Fellowship Program. L.B. is supported by a Marie Curie Individual Fellowship within the second Horizon 2020 Work Programme. J.S. acknowledges support from the Baltic American Freedom Foundation.

%\bibliography{gratingBiblio3}
%

\section{Contributions}
A.J., D.B., and V.M.A. conceived the idea. P.K., A.J., and V.M.A. planned and carried out the experiments and analysis in V.M.A.'s lab at UNM.  N.M., I.F., and F.M.B. fabricated and characterized the nanograting sensors under supervision of A.N., S.R.J.B., and V.M.A. F.M.B. assembled the optical NMR apparatus and P.K. developed the software automation. A.L., J.S., and L.B. contributed to data collection and analysis. All authors discussed results and contributed to the writing of the manuscript.

\widetext
\clearpage

\begin{center}
\textbf{\large Supplemental Materials: Solution nuclear magnetic resonance spectroscopy on a nanostructured diamond chip}
\end{center}
\setcounter{equation}{0}
\setcounter{figure}{0}
\setcounter{table}{0}
\setcounter{page}{1}
\makeatletter
\renewcommand{\thetable}{S\arabic{table}}
\renewcommand{\theequation}{S\arabic{equation}}
\renewcommand{\thefigure}{S\arabic{figure}}
\renewcommand{\bibnumfmt}[1]{[S#1]}
\renewcommand{\citenumfont}[1]{S#1}

\section{Fabrication details}
Nanogratings were fabricated on the surfaces of electronic-grade diamond chips purchased from Element 6 (initial dimensions $2\times2\times0.5$ mm$^3$, 1.1\% $^{13}$C abundance, [100]-polished faces, [110] sides). After cleaning the chips for 7 hours in a 1:1:1.3 mixture of nitric:perchloric:sulfuric acids at 200 $^{\circ}$C (from now on referred to as triacid cleaning),  each diamond chip was mounted flat on a silicon substrate. The chip was surrounded by four 0.5-mm-thick stainless-steel sheets flush with the diamond surface to avoid edge beading during spin coating. A thin film of i-CON 16 was used as an adhesive to stick the diamond and the steel sheets to the silicon substrate because it does not outgas or reflow during baking. The next step was to spin-coat i-CON 16 as an anti-reflection coating (ARC) on top of the diamond (4000 rpm, 150 $^{\circ}$C oven bake for 5 minutes), followed by spin-coating with a UV negative photoresist (NR-7 500p, 4000 rpm, 150 $^{\circ}$C hot plate bake for 1 minute).  

Interferometric lithography \cite{S_Brueck2005,S_Xia2011} was used to produce optical standing waves with a period of $\sim$400 nm. With interferometric lithography, the entire diamond surface was covered with periodic nanostructures in a few seconds of exposure time. By varying the resist, postbake, and development conditions, the grating duty cycle can be varied by at least 20-80\%. Most of the samples used in this work had approximately 50\% duty cycle. After developing for 1 minute in RD6 developer, the un-masked regions of ARC were removed using a reactive ion etch (RIE) (10 sccm O$_2$ flow, 10~W RF power) for 1 minute. For the metal mask, 3-5 nm of Cr (used for adhesion) followed by 70 nm of gold were deposited on the masked diamond chips. Liftoff was performed by immersing the chips in a sonicating acetone bath for 10 minutes to leave only Cr/Au grating structures on the diamond surface. Next, a highly anisotropic oxygen/argon etch using inductively coupled plasma (ICP) (8 sccm Ar$_2$ flow, 16 sccm O$_2$ flow, 30 W forward RF power, 450 W ICP RF power) was applied for 40 minutes to form nanogratings up to 3 $\upmu$m deep. 

In the final etch step, there is a tradeoff between etching deeper and losing the mask. Ideally, the mask would remain thick enough ($\gtrsim10~{\rm nm}$) to block ions from penetrating the tops of the nanogratings during implantation. This is because ions incident on the nanograting tops would travel on average $\sim$6 times deeper than those incident to the sidewalls due to the large difference in angles of incidence (see below). At such a large depth these NV centers will hardly register any NMR signal, reducing the overall NMR signal contrast. We found that the 40 minutes of etching used in the chips in this work was longer than optimal, and the mask had eroded over large regions of the chips prior to implantation. We have since optimized the process to use a slightly thicker metal mask such that the mask remains largely intact after 35 minutes of etching.

After fabrication, the nanograting sidewalls were implanted with nitrogen. When possible, we left the remaining metal mask using for etching to block ions from penetrating deep into the tops of the gratings. Implantation was performed by Materials Diagnostics (Albany, NY), using a $^{15}$N$^+$ ion beam with 20-200 keV implantation energy (corresponding to $\sim$5-20 nm typical NV depth), delivering a dose of $2\times10^{13}$ - $8\times10^{13}$ $^{15}$N$^+$ ions/cm$^2$ at 4 degree implantation angles (Tab.~\ref{srimTable}). For comparison, flat chips were mounted vertically (to implant at the same angle; see Fig.~2(d) inset of main text). Following implantation, the chips were cleaned in triacid overnight, then annealed in vacuum in a multi-step annealing process (800 $^{\circ}$C for 4 hours followed by 1100 $ ^{\circ}$C for 2 hours) to form an NV layer at the nanograting sidewall surfaces. This annealing procedure was selected to give a high NV$^-$ yield while minimizing the abundance of other paramagnetic impurities \cite{S_isoyaAnneal}. After annealing, the diamond chips were again cleaned in triacid for 10 hours to remove residual graphite on the surfaces and mounted in the microscope setup. Each time we removed a diamond chip from the setup, we used the same acid washing procedure to remove immersion oil, \fom~oil, CsF/glycerol solutions, or other contamination.

\section{Stopping and Range of Ions in Matter calculations}
NV center depth profiles were estimated using the Stopping and Range of Ions in Matter (SRIM) Monte-Carlo simulation \cite{S_srimRef}. The diamond chips were modeled as a pure $^{12}$C layer with 3.5 g/cm$^3$ density and 37.5 eV atom displacement threshold energy  \cite{S_displacementEnergy}. The lattice damage threshold and surface damage threshold were set to 7.35 eV and 7.5 eV, respectively. Note that SRIM simulations do not take into account crystallographic effects such as ion channeling, and therefore could lead to an underestimation of the NV implantation depth \cite{S_toyliWeis}, but are sufficiently accurate for our purposes. 

\linespread{1.0}
\begin{table}[ht]
\begin{tabular}{|c|c|c|c|c|} \hline
Chip name & Implant energy & Implant dose (2$\times$) & Effective dose (2$\times$) & Typical depth \\
%\toprule
\hline
UNM15 (flat)    &      20 keV       & $2\times10^{13}$ cm$^{-2}$      & $1.4\times10^{12}$ cm$^{-2}$    &  5 nm   \\
UNM9 (grating)  &      20 keV       & $2\times10^{13}$ cm$^{-2}$      & $1.4\times10^{12}$ cm$^{-2}$    &  5 nm   \\
\hline
UNM12 (flat)    &      60 keV       & $5\times10^{13}$ cm$^{-2}$      & $3.5\times10^{12}$ cm$^{-2}$    &  10 nm  \\
UNM11 (grating) &      60 keV       & $5\times10^{13}$ cm$^{-2}$      & $3.5\times10^{12}$ cm$^{-2}$    &  10 nm  \\
\hline
UNM16 (flat)    &     200 keV       & $8\times10^{13}$ cm$^{-2}$      & $5.6\times10^{12}$ cm$^{-2}$    &  20 nm  \\
UNM10 (grating) &     200 keV       & $8\times10^{13}$ cm$^{-2}$      & $5.6\times10^{12}$ cm$^{-2}$    &  20 nm  \\
\hline
\end{tabular}
\caption{\label{srimTable} Implantation parameters and SRIM characteristics for diamond chips studied in this work. The doses listed in the third and fourth columns were delivered twice, once at $\theta=+4^{\circ}$ and once at $\theta=-4^{\circ}$.}
\end{table}
\linespread{1.5}

\begin{figure}[ht] \begin{center}
\includegraphics[width=1.0\textwidth]{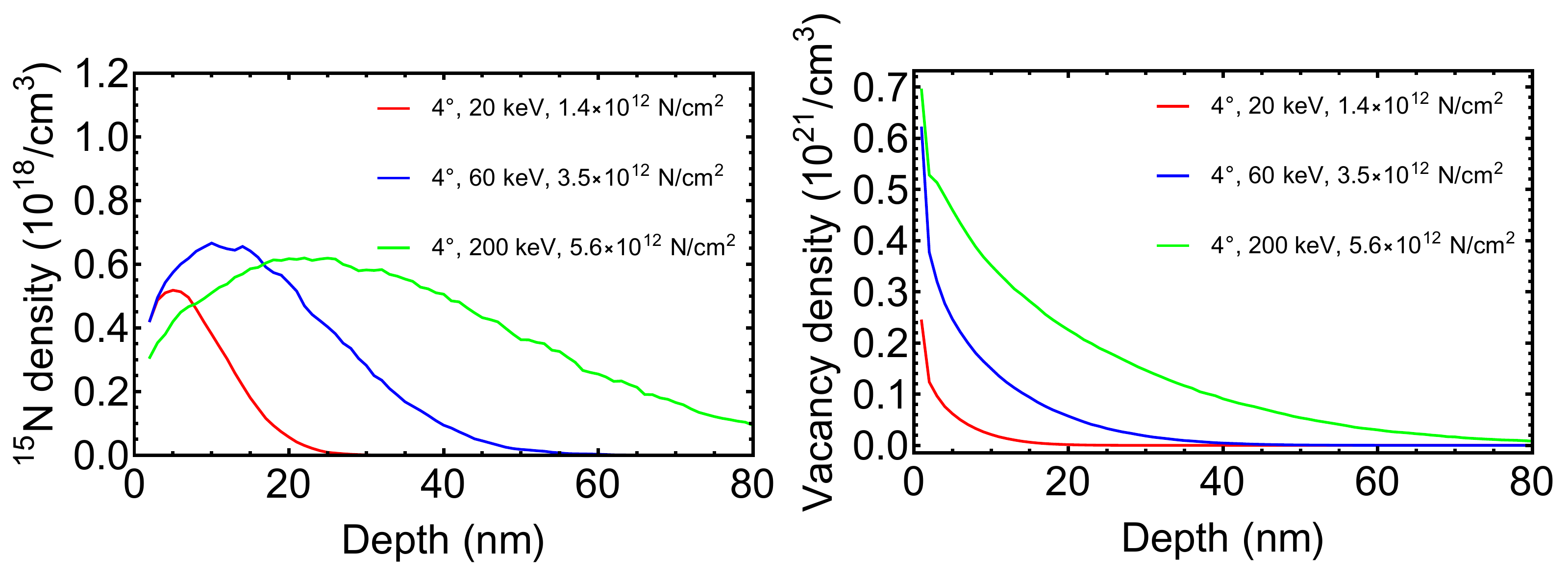}
\end{center}
\caption{\label{fig:srim}
SRIM $^{15}$N and vacancy depth profiles for our implantation conditions (Tab.~\ref{srimTable}). Legends indicate the different implantation energies and effective doses (taking into account the factor of $\sin{\theta}$ described in the text).
}
\end{figure}

Table \ref{srimTable} shows the implantation conditions studied here. One flat and one nanostructured diamond was implanted with each set of parameters, for a total of 6 chips. Both diamonds were mounted on the same substrate (as shown in Fig.~2d inset of main text) and were implanted twice, once at $\theta=+4^{\circ}$ and once at $\theta=-4^{\circ}$, where $\theta$ is the angle of incidence with respect to the substrate surface normal vector. The effective dose is the dose delivered to a single face of a flat diamond or to the sidewalls of a nanograting chip. It is calculated as the area dose (ions/cm$^2$) multiplied by $\sin{\theta}$. Doses were selected that produce a nitrogen density of $\sim10^{18}$cm$^{-3}$, which is in the range previously found optimal for NV-ensemble sensing applications \cite{S_Acosta2009,S_devience_nuclSens}. This dose still yielded less than $10^{21}$ vacancies/cm$^3$, far below the graphitization threshold ($\sim10^{22}$ vacancies/cm$^3$)  \cite{S_graphitization}, Fig. \ref{fig:srim}(right).

\linespread{1.0}
\linespread{1.5}

Figure \ref{fig:srim}(left) plots the nitrogen depth profile following implantation for different implantation conditions. Our SRIM simulations predict a $\sim$5 nm modal depth (the depth where nitrogen density is greatest) for 20 keV $^{15}$N$^+$ ion implantation, $\sim$10 nm depth for 60 keV, and $\sim$20 nm depth for 200 keV. The NV depth profiles should be similar to the simulated nitrogen profiles if there is relatively uniform nitrogen-to-NV conversion efficiency. To obtain similar NV density for all three implantation energies, we adjusted the nitrogen implantation doses to obtain approximately $0.5\times10^{18}$ N/cm$^{-3}=3~$ppm at the modal depth.
\section{Epifluorescence microscope setup}
\linespread{1.0}
\begin{figure}[ht] \begin{center}
\includegraphics[width=1.0\textwidth]{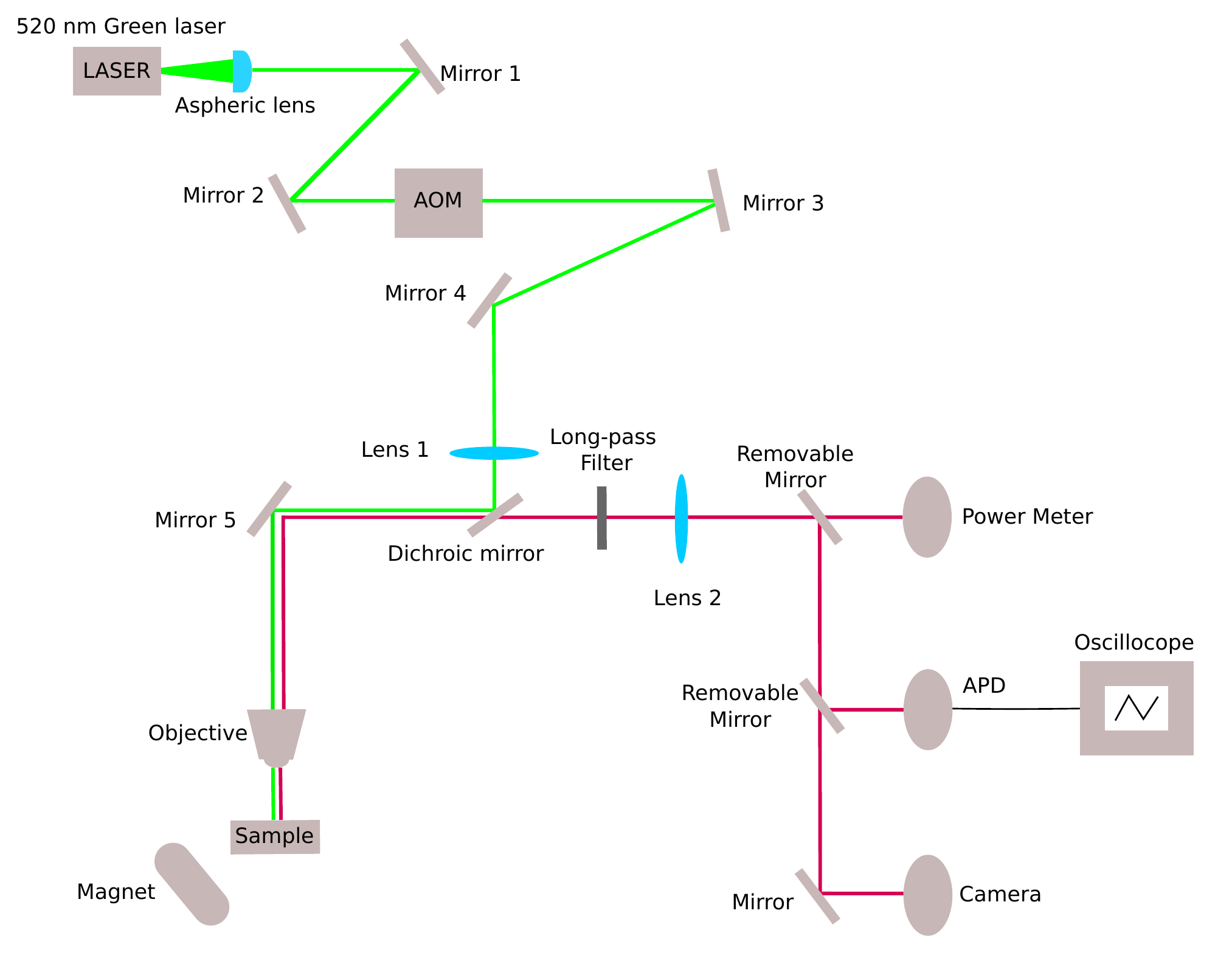}
\end{center}
\caption{\label{fig:schematic}
A schematic of the epifluorescence microscope setup. The microwave source, amplifier, switches, and acquisition computer are not shown. The computer controls the AOM and microwave switches with digital pulses and reads out averaged fluorescence time traces from the oscilloscope. Green lines indicate the 520 nm pump laser path and red lines indicate the NV fluorescence beam path.
}
\end{figure}
\linespread{1.5}
Figure \ref{fig:schematic} shows the experimental setup. Light from a 520 nm diode laser is used to pump and probe the NV centers (140 mW at the objective). The excitation beam is shaped to illuminate a (25 $\upmu$m)$^2$ patch on the diamond. The oil-immersion microscope objective (1.25 NA, 100$\times$ magnification) collects the NV fluorescence, which is detected with an avalanche photodiode (APD). A camera is used to estimate the beam spot size and image the nanogratings, and a power meter to is used to monitor excitation power and fluorescence intensity. An oscilloscope measures the APD output voltage, reporting fluorescence time traces to the experimental control computer (not shown). The computer controls a I/Q modulated microwave generator, which is modulated with fast TTL pulses using microwave switches (not shown). The microwaves are amplified and applied to the diamond with a wire loop fabricated on a glass coverslip.
\section{Diamond chip characterization}
\linespread{1.0}
\begin{table}[ht]
\begin{tabular}{|c|c|c|c|c|c|}
\hline
Chip name & Fluorescence &  Rabi contrast & $T_2$ (176 pulses) & $T_2$ (1 pulse) & $p$ exponent\\
%\toprule
\hline
UNM15 (flat)    &      4 nW         & 0.035      & 54$\pm$5 $\upmu$s    &  N/A  &  N/A \\
UNM9 (grating)  &      76 nW        & 0.027      & 53$\pm$6   $\upmu$s    &  4.1$\pm$0.5 $\upmu$s  & 0.51$\pm$0.03 \\
\hline
UNM12 (flat)    &      20 nW        & 0.035      & 83$\pm$5 $\upmu$s      &  6.6$\pm$0.5 $\upmu$s  & 0.49$\pm$0.02 \\
UNM11 (grating) &      580 nW       & 0.027      & 75$\pm$9 $\upmu$s      &  3.2$\pm$0.5 $\upmu$s  & 0.63$\pm$0.04 \\
\hline
UNM16 (flat)    &      15 nW        & 0.019      & 73$\pm$8 $\upmu$s      &  2.9$\pm$0.7 $\upmu$s  & 0.63$\pm$0.06 \\
UNM10 (grating) &      795 nW       & $\gtrsim0.014$      & 70$\pm$3 $\upmu$s      &  2.7$\pm$0.5 $\upmu$s  & 0.62$\pm$0.04 \\
\hline
\end{tabular}
\caption{\label{samplesTable} Chip characterization done with the epifluorescence microscope at $B_0\approx5~{\rm mT}$ aligned along one of the N-V axes. Here we list the fluorescence intensity when illuminated by 140 mW of 520 nm laser light collected through a 1.25 NA oil objective lens. The contrast represents the fluorescence contrast of ($\sim$15 MHz) Rabi oscillations between the $m=0$ and $m=-1$ states. The NV $T_2$ is evaluated for an XY8-22 pulse sequence (176 $\pi$-pulses). $T_2$ increases with more $\pi$-pulses with a power law dependence with exponent $p$.}
\end{table}
\linespread{1.5}

\linespread{1.0}
\begin{figure*}[ht]
\begin{center}
\includegraphics[width=1.0\textwidth]{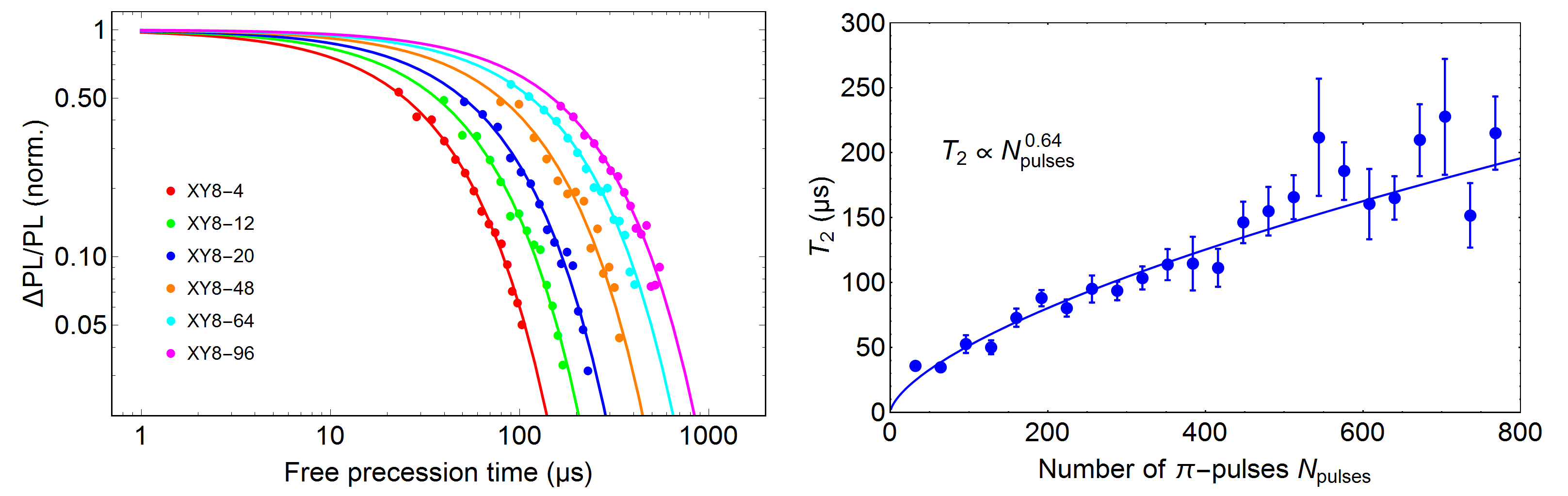}
\end{center}	
\caption{\label{XY8Nfig}
(Left) XY8-N $T_2$ measurements for diamond chip UNM6 (a nanograting chip not included in the above tables; 33\%:67\% diamond/gap duty cycle 40 keV implant, $5\times10^{13}~^{15}$N$^+$/cm$^2$, $\pm 5^{\circ}$ degree implant angle). The plotted fit functions are exponential decays used to extract $T_2$. (Right) Adding more $\pi$-pulses to the pulse sequence improves $T_2$ with a power law scaling.
}
\end{figure*}
\linespread{1.5}
As described in the main text, we demonstrate that the nanograting chips have brighter fluorescence without sacrificing NV coherence or contrast. Table \ref{samplesTable} lists the results of Fig.~2(d)-(e) in greater detail. Figure \ref{XY8Nfig} shows the results of typical NV XY8-N experiments used to characterize the NV $T_2$ coherence. Note that the NMR experiments performed in this work required only a modest $T_2$ in the tens of $\upmu$s range. This is because the maximum phase accumulation time is already restricted to this range due to rapid molecular diffusion within the analyte. As such, we were able to avoid using $^{12}$C-enriched diamond layers and implant with a high $^{15}$N$^+$ density even though the magnetic noise from $^{13}$C nuclei and paramagnetic nitrogen defects in the diamond tends to reduce the NV $T_2$ coherence time.
\linespread{1.0}
\begin{figure*}[ht]
\begin{center}
\includegraphics[width=0.8\textwidth]{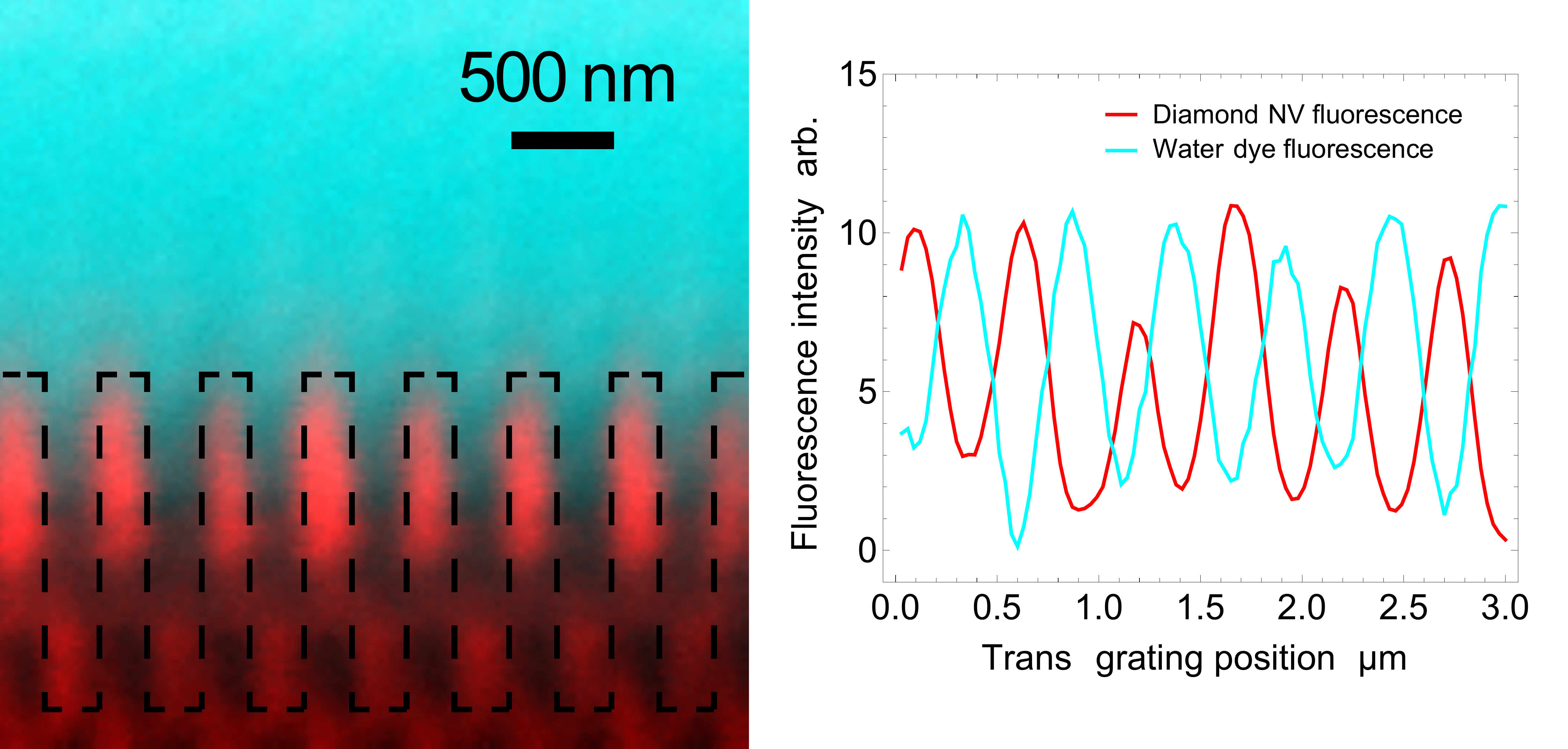}
\end{center}	
\caption{\label{wettingAntiCorr}
(Left) Confocal microscopy image from Fig.~2(c) in the main text. (Right) Line cut of fluorescence across the grating shows that fluorescence from water dye is anticorrelated with the fluorescence from diamond, confirming wetting.
}
\end{figure*}

The nanograting chips can only exhibit improved NMR sensitivity if the analyte solution wets the diamond surface. NV centers from regions that do not wet are too far from analyte to register its NMR spectrum and thus reduce the overall signal contrast. Wetting can be a challenge with dense, high-aspect-ratio nanostructures, often depending on factors such as surface termination \cite{S_Xia2008, S_Xia2010}. To confirm that water wets the grating surfaces, we immersed a grating chip in an Alexa 405 dye/water solution and measured NV and dye fluorescence with a confocal microscope Leica TCS SP8, Fig.~\ref{wettingAntiCorr}(left). Figure \ref{wettingAntiCorr}(right) shows that dye and NV fluorescence (when illuminated with blue and green light, respectively) are spatially anticorrelated. This confirms that the nanograting sidewalls are largely in contact with the analyte. The improved NMR sensitivity exhibited by nanograting chips reported in the manuscript is further confirmation, as this is also only possible if there is substantial wetting of the nanograting sidewalls.

\section{Experimental procedure}
After mounting a diamond chip in the microscope, we align the $B_0$ field from a permanent magnet by measuring the optically-detected magnetic resonance (ODMR) frequencies of the four NV orientations. Good field alignment is important because $T_2$ is maximized and the NV contrast is maximized when $B_0$ is aligned along the N-V axis \cite{S_stanwix}, resulting in better sensitivity. We use the NV sub-ensemble aligned with $B_0$ for correlation spectroscopy, while the other three NV sub-ensembles do not participate in the measurement and contribute background fluorescence. After aligning, we calculate the $B_0$ magnitude from the ODMR spectrum.  We position the microwave wire to achieve a reasonably-fast Rabi frequency ($10\mbox{-}15~$MHz) and test the correlation spectroscopy experiment with the AC magnetic field from a calibrated test coil (driven by a sine wave from a function generator). 

When performing NMR spectroscopy, we select parameters for the correlation pulse sequence based on the following principles. Laser light pulses have 5 $\rm \upmu s$ duration. The first $\sim1~{\rm \upmu s}$ is used for readout, while the remainder is used to efficiently repolarize the NV centers. The separation between $\pi$ pulses, $2\tau$, is chosen to match half the nuclear Larmor period of the target spin, $4\tau=\tau_L$. We select the number of repetitions (the ``N'' in XY8-N) which produces the highest signal-to-noise-ratio (SNR), at constant measurement time, using protons in pure glycerol as a convenient test sample. The step size between $\tilde{\tau}$ values is chosen to be $\sim\tau_L/4$, and the longest value is chosen to match the approximate duration of the XY8-N sequences. This represents a compromise between obtaining high SNR (which would seek to minimize the overall measurement time) and obtaining high spectral resolution (which would prefer to use as long a $\tilde{\tau}$ as possible). In every experiment we alternate the phase of the final $\pi/2$ pulse and subtract the fluorescence signals, resulting in fast common-mode rejection of fluorescence intensity drifts. 

We have noticed that the fluorescence intensity depends on the time between laser pulses in a non-trivial manner, presumably due to complex NV$^0$ / NV$^-$ dynamics. To circumvent related systematic effects, we add buffer time between the last microwave $\pi/2$-pulse and the laser readout pulse to ensure that the time between laser pulses remains constant while we sweep $\tilde{\tau}$. However, this means most experiments take $\sim1.5\times$ longer than they should because of the buffer time. We average fluorescence readout time traces on an oscilloscope, which introduces additional dead time. The oscilloscope misses triggers while averaging and data processing, and it also spends time transferring averaged time traces to the acquisition computer. These dead times collectively make the actual experiments roughly $\sim2.5\times$ slower than necessary, which we will improve in future setups. When determining $t_{avg}$ in the main text, we neglect this dead time. 

\section{Molecular diffusion analysis}
\begin{figure}[ht]
\begin{center}
\includegraphics[width=0.4\textwidth]{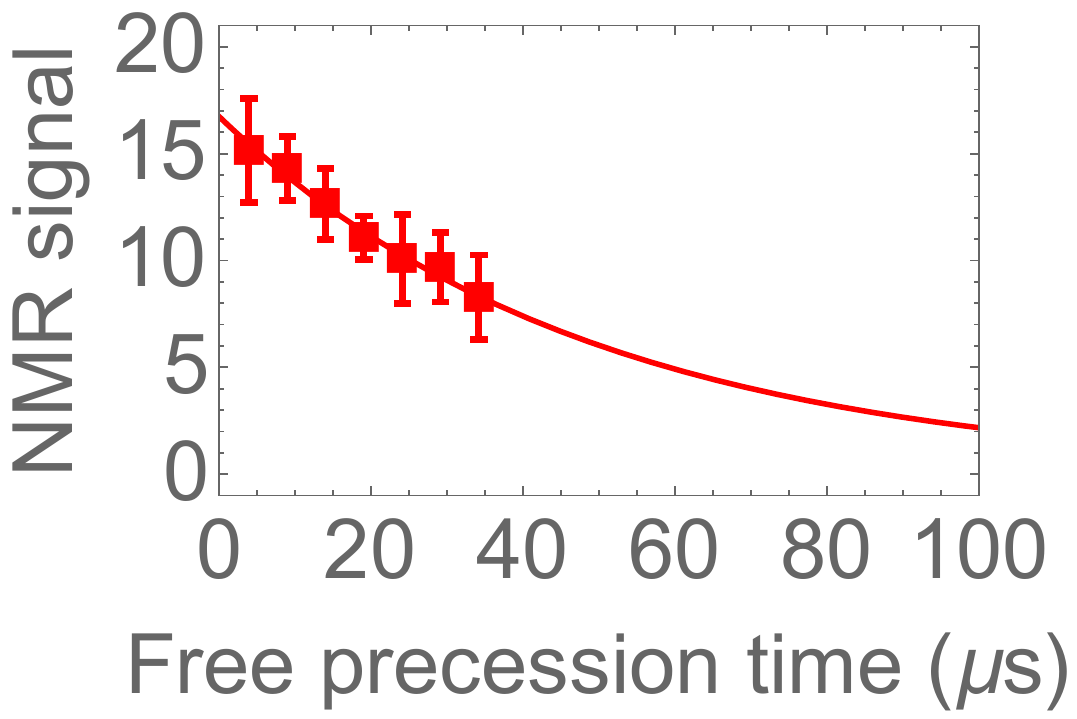}
\end{center}	
\caption{\label{diffExample}
A glycerol molecular diffusion NMR decay curve measured using the technique described in this text. The sensor was UNM11 (60 keV grating chip). 
}
\end{figure}

To determine the effects of molecular diffusion, we acquired $^1$H correlation data from all three nanograting chips with two different analytes, glycerol and Olympus Type-F microscope immersion oil. For the 200 keV chip (UNM10) with glycerol the correlation signal was too weak to properly analyze. We binned the NV correlation spectroscopy measurement values into different sets of ranges of $\tilde{\tau}$ and obtained the signal amplitude from each bin from its Fourier transform. We fit the signal amplitude versus time with a decaying exponential to extract the nuclear-spin correlation time $\tau_C$ (examples shown in Fig.~\ref{diffExample}). We repeated using different binning lengths/techniques to obtain the approximate uncertainties $\sigma_n$ for the error bars shown in Fig.~6(c) of the main text.

For fitting to a molecular diffusion model, we simulated 20, 40, 60, 80, and 200 keV implantation with the SRIM calculation described above. We extracted the modal implantation depth for each energy, and fit these data with a spline interpolation function to generate a smooth function, $f(E)$, that maps energy $E$ to depth, $d_{NV} = f(E) \times E$. We then fit the $\tau_C$ versus implantation energy data with the function $\tau_C = d_{NV}^2/D = f(E)^2 \times E^2 / D$, revealing the diffusion coefficient, $D$. We used a weighted fit (with the $n$th point in the fit weighted proportional to $1 / \sigma_n^2$) because the 20 keV correlation times were extracted with small relative uncertainties while the 200 keV correlation measurements were noisier.

\linespread{1.0}
\linespread{1.5}

\section{Thermal and statistical polarization comparison}
In the case where $h \gamma_{nucl} B_0 \ll k_B T$ (the Curie's law high-temperature limit), where $B_0$ is the applied field, $k_B = 1.381 \times 10^{-23}$ J/K is the Boltzmann constant, and $T$  is the temperature, the field from nuclear thermal polarization is (ignoring numerical factors of order 1 from the discrete nuclear spatial distribution):
\begin{equation}
B_{therm} \sim \frac{\mu_0}{4 \pi} \frac{ h^2 \gamma_{nucl}^2 B_0 \rho}{4 k_B T},
\end{equation}
where $\mu_0 = 4\pi \times 10^{-7}$ m$\cdot$T/A is the vacuum permeability, $h = 6.626 \times 10^{-34}$ J$\cdot$s is the Planck constant, $\gamma_{nucl}$ is the nuclear gyromagnetic ratio in MHz/T (42.58 MHz/T for $^1$H), and $\rho$ is the nuclear spin number density. This field is roughly constant for a distance of order the thickness of the analyte after which it falls off as the distance cubed.

In contrast, NV centers near the surface of a flat diamond surface experience a magnetic field from the statistical polarization of the nuclei. Averaged over time (and over the NV ensemble), the field's magnitude approaches the standard deviation of the normally-distributed random magnetization from configurations of spins within each NV center's detection volume. The field component along the N-V axis from the nuclear statistical polarization, $B_{RMS}$, is given by \cite{S_linhNMRdepth}:
\begin{equation}  \label{BrmsEqn}
B_{RMS}^2 =  \frac{\pi (8 - 3 \sin(\alpha)^4)}{128}  \left( \frac{\mu_0 h \gamma_{nucl}}{4 \pi}  \right)^2 \frac{\rho}{d_{NV}^3} 
= P(\alpha)  \left( \frac{\mu_0 h \gamma_{nucl}}{4 \pi}  \right)^2 \frac{\rho}{d_{NV}^3},
\end{equation}
where $d_{NV}$ is the NV depth below the diamond surface. The function $P(\alpha) = \frac{\pi (8 - 3 \sin(\alpha)^4)}{128}$ is a geometric factor that comes from the angle $\alpha$ the N-V axis makes with the normal to the substrate surface (see Fig.~6 in Ref.~\cite{S_linhNMRdepth}). Some previous experiments used flat diamond chips polished along the [100] crystallographic direction. The flat samples in the present manuscript also use this geometry. In this case, all four NV orientations have the same $\alpha=\cos^{-1}(\sqrt{1/3})=54.7^{\circ}$ angle to the surface. The grating sidewalls in our nanograting chips are approximately along the [110] crystallographic direction. Two of the NV orientations lie in-plane (with slightly worse sensitivity than with [100]) and the other two orientations lie out-of-plane (with slightly better sensitivity). Table \ref{pointingDirections} details $P(\alpha)$ for these geometries. 

\linespread{1.0}
\begin{table}[ht]
\begin{tabular}{|c|c|c|c|}
\hline
Diamond geometry & Surface	 & $\sin(\alpha)$ & $P(\alpha)$ \\
%\toprule
\hline
Flat diamond                & [100] &  $\sqrt{\frac{2}{3}}$ & $\frac{5 \pi}{96} \approx 0.052 \pi$ \\ \hline
Grating (out-of-plane)      & [110] &  $\sqrt{\frac{1}{3}}$ & $\frac{23 \pi}{384} \approx 0.060 \pi$ \\  \hline
Grating (in-plane)          & [110] &  $1$                  & $\frac{5 \pi}{128} \approx 0.039 \pi$\\ \hline
\end{tabular}
\caption{\label{pointingDirections} A table listing $P(\alpha)$ for relevant NV pointing directions, which give rise to different $B_{RMS}^2$. The NVs in our nanograting sidewall surfaces have two angles with respect to the sidewall normal direction, one of which is $\sim$50\% more sensitive to the nuclear statistical polarization than the other. }
\linespread{1.5}
\end{table}

\linespread{1.0}
\begin{figure}[ht]
\begin{center}
\includegraphics[width=1.0\textwidth]{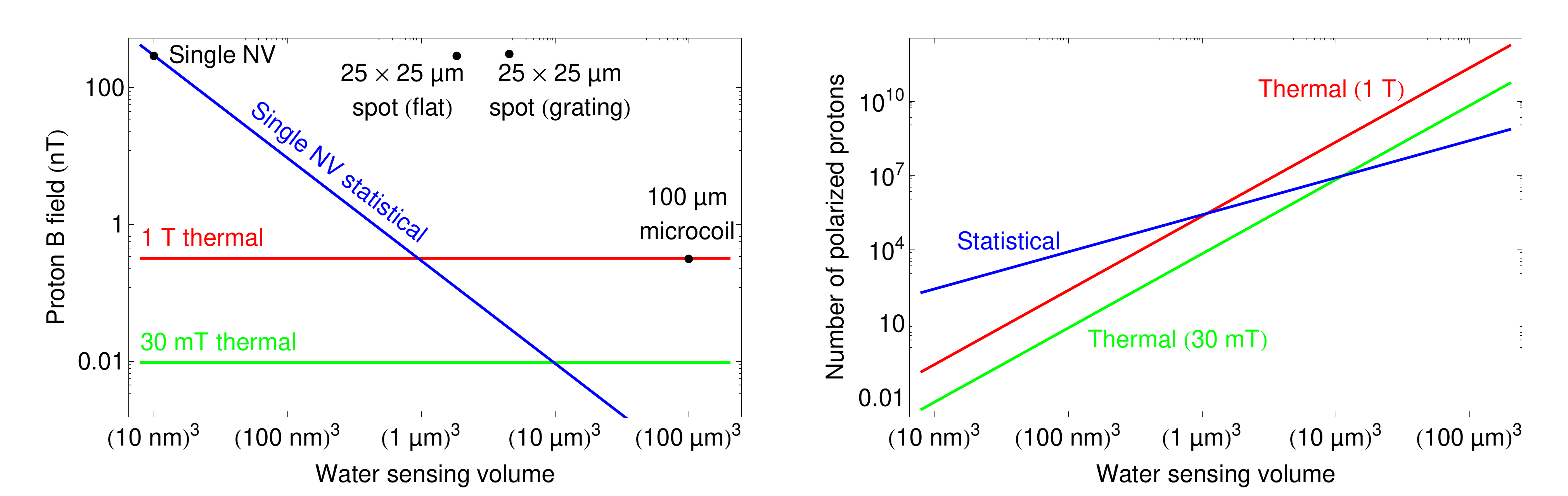}
\end{center}	
\caption{\label{statTherm}
Comparison between thermal and statistical nuclear polarization as a function of detection volume for $^1$H nuclei in water.
(Left) The typical magnetic field from thermally-polarized nuclei (red and green lines) depends linearly on $B_0$ but is independent of sample volume. The statistical magnetic field is independent of $B_0$, but for a single NV (blue line) it falls off rapidly with detection volume since that requires a greater NV depth. Using an NV ensemble with a constant depth increases the sensing volume without sacrificing the field strength. Here we plot the expected $B_{RMS}$ for 10 nm deep NV ensembles in a (25 $\upmu$m)$^2$ spot for flat and grating chips. The volume for a flat chip is not practical for fluids, as it is only 10 nm thick. However the volume for the nanogratings case is realistic; it only assumes the nanograting grooves are filled to the top as in the present experiments.
(Right) The number of thermally-polarized nuclei increases linearly with $V$ while the number of statistically-polarized nuclei increases like $\sqrt{V}$.
}
\end{figure}
\linespread{1.5}
Figure \ref{statTherm}(left) compares the $^1$H statistical and thermal fields as a function of water volume, $V$. For thermal polarization, the field is independent of sample volume provided that the NV centers are located sufficiently close ($\lesssim V^{1/3}$). For statistical polarization, the field is a strong function on the NV depth, $d_{NV}$. Intuitively this is because the technique relies on the random net difference in magnetization expressed as a fraction. For small ensembles of nuclei this fraction is much larger, producing a correspondingly larger magnetic field. Assuming the NV statistical sensing volume is $V\approx d_{NV}^3$, the field scales as $V^{-1/2}$.

Figure \ref{statTherm}(right) compares the number of polarized spins which contribute to an NMR signal for thermal and statistical polarization detection. For spin-1/2 nuclei in a detection volume $V$, there are $N_{therm} = \frac{h \gamma_{nucl} B_0 }{2 k_B T} \rho V$ thermally-polarized nuclei and $N_{stat} = \sqrt{\rho V}$ statistically-polarized nuclei  \cite{S_herzog_thermVsStat}. For sufficiently small $B_0$ and $V$, the statistical polarization dominates over thermal. An additional advantage is the correlation spectroscopy technique used here does not require RF pulses to control the nuclei, avoiding the need for high RF power to manipulate nuclear spins and allowing for a broader NMR spectral bandwidth. Moreover, since the nuclei are never manipulated (and are therefore always in equilibrium), there is never a need to wait several nuclear longitudinal spin relaxation times $T_1$ before starting a new experiment. This is particularly advantageous if the nuclear $T_1$ is much longer than the nuclear transverse spin relaxation time, $T_1>>T_2$, where a duty cycle of order $T_1/T_2$ would be required in conventional NMR.

\section{Theoretical nuclear concentration sensitivity}
In correlation spectroscopy, the nuclear Larmor precession is encoded as oscillations in the NV spin populations and resulting fluorescence intensity. Following the analysis in Ref.~\cite{S_c13CorrSpec} for correlation spectroscopy using Hahn echo, we treat the nuclear statistical magnetization as an AC magnetic field (frequency $f_{nucl}$) with a random amplitude that is normally distributed with a variance $B_{RMS}^2$. The correlation time of the nuclear spins is presumed to be sufficiently long that the nuclear field is constant throughout a single pulse sequence but short enough that all possible amplitudes are averaged over many experiments. Assuming the NV centers experience a small statistical nuclear field, $B_{RMS} \ll (4 \gamma_{NV}\tau_L)^{-1}$, the NV fluorescence intensity after a correlation spectroscopy pulse sequence is:
\begin{equation}
 F= N_{phot} \left( 1 - \frac{C}{2} \right) + \frac{ N_{phot} C}{2} \times\frac{1}{2} \left[  \frac{4 \gamma_{NV} B_{RMS} }{ f_{nucl} }  \sin^2 \left( \frac{ 2 \pi f_{nucl} \tau }{2} \right) \right]^2 \cos \left( 2 \pi f_{nucl} (2 \tau + \tilde{\tau} ) \right).
\end{equation}
Here $N_{phot}$ is the number of photons collected in one readout, $C$ is the maximum possible fluorescence contrast in the strong $B_{RMS}$ limit, $\gamma_{NV}=28.03~$GHz/T is the NV gyromagnetic ratio, $f_{nucl} = \gamma_{nucl} B_0 $ is the nuclear Larmor frequency, $\tau$ is the time duration between the NV $\pi / 2$-pulse and $\pi$-pulse, and $\tilde{\tau}$ is the free precession time between the two Hahn echo sequences. In a Hahn-echo correlation sequence, we tune the pulse spacing to match the nuclear Larmor frequency such that $2\tau=\tau_L$ and replace $f_{nucl}$ with $1/T_{tot}$, where $T_{tot}$ is the total phase accumulation time. This simplifies the expression for fluorescence intensity to
\begin{equation}
    F = N_{phot} \left( 1 - \frac{C}{2} \right) + 4 N_{phot} C \left(  \gamma_{NV} B_{RMS} T_{tot}  \right)^2 \cos \left( 2 \pi f_{nucl}  \tilde{\tau}  \right).
\end{equation}
This expression also holds for the more complicated XY8-N sequence. The primary difference is the phase accumulation time, $T_{tot}$, is a factor 4N longer in the XY8-N sequence as compared to the Hahn echo version.

If the measurement is photon shot-noise-limited, then $\Delta F = \sqrt{N_{phot}} (1 - C/2) \approx \sqrt{N_{phot}}$, and the minimum detectable field strength (SNR=1) is given by: 
\begin{equation} \label{NVdeltaB2Eqn}
    \Delta B_{RMS,min}^2 = \frac{\Delta F}{| \partial F / \partial B_{RMS}^2 | } = \frac{1 }{4 \sqrt{ N_{phot} } C ( \gamma_{NV} T_{tot}  )^2 }.
\end{equation}
Combining Eq.~\ref{BrmsEqn} and Eq.~\ref{NVdeltaB2Eqn} yields $\rho_{min}$, the minimum-detectable magnetic spin concentration, after one pulse sequence:
\begin{equation} \label{rhoMinEqn}
    \rho_{min}(\mathrm{SNR=1}) = \frac{1}{P(\alpha)(\mu_0 \hbar \gamma_{NV}\gamma_{nucl})^2}\times \frac{d_{NV}^3}{T_{tot}^2C\sqrt{N_{phot}}}.
\end{equation}
To evaluate the minimum-detectable concentration after one second of averaging, we divide by $\sqrt{N_r}$, where $N_r$ is the number of readouts per second. We also replace $N_{phot}$ with the product of the number of NVs ($N_{NV}$) and the number of photons collected per NV per readout ($\eta$). Imposing the NMR standard requirement of SNR = 3 in 1 s, we obtain (after substituting $h = 2\pi \hbar$):
\begin{equation}\label{eq:rhomin3}
     \rho_{min}( \mathrm{SNR=3~in~1~s} ) = \frac{3}{P(\alpha)(\mu_0 \hbar \gamma_{NV}\gamma_{nucl})^2} 
     \times \frac{d_{NV}^3}{T_{tot}^2C\sqrt{\eta N_{NV} N_r}} .
\end{equation}

In Fig.~1(a) of the main text we plot $\rho_{min}( \mathrm{SNR=3~in~1~s} )$ as a function of analyte volume using the following parameters: $\gamma_{nucl}=40.08~{\rm MHz/T}$, $P(\alpha)=23\pi/384$ (Tab. \ref{pointingDirections}), $d_{NV}=5~{\rm nm}$, $T_{tot}$ = 25~$\upmu$s, $C=0.02$, $N_r=2\times10^{4}$, and $\eta=0.03$. To determine how $N_{NV}$ scales with volume, we assume that a dose of $\rm 5\times10^{13}~^{15}N^+/cm^2$ is delivered to the chip, and, after annealing, the nitrogen-to-NV conversion efficiency \cite{S_Acosta2009} is $10\%$, resulting in $\rm 5\times10^{4}~NV^-/{\upmu m}^2$. Thus $N_{NV}$ scales linearly with sensor area. We then assume the gratings are 2 ${\rm \upmu m}$ tall with $50\%$ duty cycle and that analyte fills the gratings flush to the top (i.e. there is no gap between coverslip and nanogratings). In this case, the analyte volume also scales linearly with sensor area. Combining with Eq.~\eqref{eq:rhomin3}, we find that the minimum detectable concentration $\rho_{min}\propto{\rm volume}^{-1/2}$ as seen in Fig. 1(a) of the main text. For 1 pL of analyte, $N_{NV}\approx5\times10^{7}$; for 1 nL of analyte, $N_{NV}=5\times10^{10}$, etc. 

Using these theoretical values, we determine $\rho_{min,theory}\approx4\times10^{22}~{\rm spins/L}$ for the $\sim 1~{\rm pL}$ analyte volume used in this work. This concentration sensitivity represents a giant improvement over previous small-volume NMR works; it is comparable to microslot NMR at high field, but with a 4-orders-of-magnitude smaller volume. While our experimental values for $\rho_{mim}$ still represent a large improvement over previous work, they are about 150 times larger than this theoretical estimate. The experimental parameters are consistent with $T_{tot}\approx 25~{\rm \upmu s}$ and $N_r\approx2\times10^{4}$ (when dead time is excluded), but the experimental fluorescence levels ($\propto\eta N_{NV}$) are about $5\times$ lower than expected from the parameters in the calculation, likely due to suboptimal N-to-NV conversion efficiency. An even larger deviation between experiment and theory is the experimental contrast, $C\approx0.004$, which is approximately 5 times lower than ideal. We attribute this to pulse errors, as $C$ decreases when descreasing the Rabi frequency or increasing the number of pulses used in the sequence \cite{S_nirDDarb}. Another deviation from theory is the $\sim2.5\times$ reduction in signal between flat and nanostructured samples discussed in the text (presumably due to deep implantation of nanograting tops/bottoms and imperfect wetting).  

The above three factors account for $\sim28\times$ difference in $\rho_{min}$ between theory and experiment. The remaining factor of $\sim5$ appears to be due to an over-estimation of the expected field strength from the analyte. From Eq.~\eqref{BrmsEqn}, the field strength for Fomblin is given by $B_{RMS,theory}^2=0.37~{\rm \upmu T}^2$ at an NV depth $d_{NV}=5$ nm. This is $\sim9$ times larger than the observed value in Fig.~4(a) in the main text, and largely accounts for the remaining observed discrepancy in $\rho_{min}$. This over-estimation suggests the chips have a deeper characteritic NV depth than expected. Possible reasons include a deviation in NV depth profile from SRIM, the presence of a few-nm surface layer of water or hydrocarbons, and/or the presence of debris or other interface issues that reduce the overall sensor-sample contact. If instead we use $d_{NV}=10$ nm, we find $B_{RMS,theory}^2=0.04~{\rm \upmu T}^2$, identical to the observed value in Fig.~4(a). Note that if the characteristic NV depth is $\sim2\times$ deeper than expected, it would reduce the fitted diffusion values in Fig. 6 of the main text by a factor of $\sim4$. In future work we will explore lower-energy implantation to reduce $d_{NV}$.

\section{Concentration sensitivity assessment}
Before settling on CsF in glycerol, we considered several analytes and target spin species to characterize our spin concentration sensitivity $\rho_{min}$. A major requirement was that the analyte have a sparser concentration than those used in previous NV and picoliter NMR work. For example, it should be smaller than the proton density in water, which is $6.7\times10^{25}$ spins/L = 110 M). Table \ref{solTable} lists  example nuclear densities (top half) and some candidate solutions (bottom half). Ideally we want a viscous solvent with slow molecular diffusion $D$ to maximize the nuclear correlation time $\tau_C$. Glycerol (C$_3$H$_8$O$_3$) is a promising solvent choice because it is viscous and is a common solvent, dissolving many different molecules. Although $^1$H is a desirable nucleus to detect ($\gamma_{nucl}$ is large and many chemicals contain hydrogen), we avoided using it because diamond surfaces can have a $\sim$1 nm adsorbed water or hydrocarbon layer, which produces a large background proton signal even in the absence of analyte \cite{S_rugar2015}. Furthermore, the baseline $^1$H concentration in the commonly-available deuterated glycerol (glycerol-d$_8$, 98\% deuterated) is comparable to or larger than the saturation concentration of common analytes like sucrose ($\rm C_{12}H_{22}O_{11}$). We therefore restricted our search to analytes featuring NMR-active nuclei other than protons.

The bottom half of Tab.~\ref{solTable} considers alternative solutions and target nuclei. The figure of merit is $B_{rms}^2 \propto \rho \gamma_{nucl}^2$ from Eq.~\ref{BrmsEqn}, so we considered nuclei with large $\gamma_{nucl}$. Although $B_{RMS}^2$ for saturated solutions of Na$_2$CO$_3$ is slightly larger than for CsF, $^{23}$Na is spectrally difficult to distinguish from $^{13}$C internal to the naturally-abundant diamond chips. For NMR peaks to be distinguishable their linewidth must be smaller than $1/\tau_C\approx100~{\rm kHz}$. For peaks due to $^{23}$Na and $^{13}$C, this becomes feasible only at $B_0\gtrsim180~{\rm mT}$, which was out of the range studied here. Thus, we chose CsF in glycerol.  The CsF saturation concentration was unavailable in the literature, but we determined it to be between 20-40\% by weight by varying the CsF concentration and monitoring precipitate levels.
\linespread{1.0}
\begin{table}[ht]
\begin{tabular}{|c|c|c|c|c|}
\hline
Spin/analyte & $\gamma_{nucl}$ (MHz/T) & Abundance & $\rho$ (spins/L)  & Comments \\
%\toprule
\hline
$^1$H in H$_2$O & 42.58 &  100\% &   $6.7\times10^{25}$   &  \\
$^1$H in IMMOIL-F30CC oil & 42.58  &  100\% &   $6\times10^{25}$   & \cite{S_loretzAPL} \\
$^1$H in glycerol & 42.58  &  100\% &   $6.6\times10^{25}$   &   \\
$^{13}$C in glycerol & 10.71  &  1.1\% &   $2.7\times10^{23}$   &   \\ 
$^1$H in 98\% glycerol-d$_8$ & 42.58  &  2\% &   $1.4\times10^{24}$   & Sigma-Aldrich 447498   \\ 
$^1$H; saturated sucrose in 98\% glycerol-d$_8$ & 42.58  &  100\% &  $2.6\times10^{24}$ & Ignores $^1$H from above line \cite{S_sucroseInGlycerol} \\
$^{19}$F in \fom & 40.08  &  100\% & $4.0\times10^{25}$   & Sigma-Aldrich 317993 \\ 
$^{13}$C in diamond crystal & 10.71  &  1.1\% &   $1.9\times10^{24}$   & Spins fixed in crystal lattice \\ 

\hline
$^{19}$F; 0.2 g CsF in 1 g glycerol &  40.08  &  100\% &  $1.0\times10^{24}$ & What we used ($\star$) \\
$^{23}$Na; saturated NaCl in glycerol & 11.27  &  100\% &  $1.1\times10^{24}$ & 12$\times$ weaker $B_{RMS}^2$ than $\star$ \cite{S_naclInGlycerol}\\
$^{23}$Na; saturated Na$_2$CO$_3$ in glycerol & 11.27  &  100\% &  $1.4\times10^{25}$ & 1.1$\times$ stronger $B_{RMS}^2$ than $\star$ \cite{S_glycerolTable} \\

$^{11}$B; saturated B(OH)$_3$ in glycerol & 13.66  &  80.1\% &  $3.0\times10^{24}$ & 3.5$\times$ weaker $B_{RMS}^2$ than $\star$ \cite{S_glycerolTable} \\

\hline

\end{tabular}
\caption{\label{solTable} Comparison of analytes with different nuclear spin densities. The top half includes relevant reference nuclear densities, and the bottom half lists solutions we considered for testing. Note that the solution volume is slightly larger than the original solvent volume after adding a solute, but for simplicity we neglect this typically small effect. We also assume that the solute molecular diffusion constant $D$ is the same as the self-diffusion of glycerol, and that the diffusion is unaffected by dissolving solutes.
}
\end{table}
\linespread{1.5}
\section{Optimal NV layer depth}
Equation \eqref{eq:rhomin3} provides insight on how to optimize the NV layer depth, $d_{NV}$, to achieve the smallest detectable $\rho_{min}$ in the limit of fast molecular diffusion and shallow NVs. According to Eq.~\eqref{eq:rhomin3}, in a correlation spectroscopy measurement, $\rho_{min}\propto d_{NV}^3/(T_{tot}^2\sqrt{N_r})$. If the XY8-N sequences dominate the correlation experiment duration, then $N_r\propto 1/T_{tot}$ since fewer experiments can be done in 1 second if the duration is longer. Thus, $\rho_{min} \propto d_{NV}^3 / T_{tot}^{3/2}$. If the molecular diffusion is fast compared to the NV $T_2$, then we choose $T_{tot} \approx \tau_C = 2 d_{NV}^2 / D$ \cite{S_linhNMRdepth}. Substituting this expression in the previous one, we  conclude that $\rho_{min}$ is independent of $d_{NV}$. Intuitively, this is because a deeper NV feels a weaker $B_{RMS}^2$ but can compensate by querying it for longer before the nuclear field randomizes.

\section{Comparing $\rho_{min}$ specifications in Fig.~1}
\linespread{1}
\begin{table}[]
\centering
\begin{tabular}{|c|c|c|c|c|c|c|c|}
\hline
Technique & Ref. & Analyte & $B_0$ field & \begin{tabular}[c]{@{}c@{}}RF pulses\\ for nuclei\end{tabular} & Volume  & \begin{tabular}[c]{@{}c@{}}Number\\ of spins\end{tabular} & Detection    \\

\hline

\begin{tabular}[c]{@{}c@{}}Cryogenic\\ probe\end{tabular}     & \cite{S_kovacsCryoProbeReview}   & \begin{tabular}[c]{@{}c@{}}$^1$H in 2 mM of\\ sucrose in D$_2$O\end{tabular}   & 14 T & yes  & 30 $\upmu$L   & \begin{tabular}[c]{@{}c@{}}$6\times10^{14}$\\ (1 nmol)\end{tabular}   & inductive    \\
\hline

\begin{tabular}[c]{@{}c@{}}Atomic\\ magnetometer\end{tabular} & \cite{S_micah2008}   & $^1$H in water & \begin{tabular}[c]{@{}c@{}}0.7 T prepolarize\\ 0 T detection\end{tabular} & yes & 1 $\upmu$L    & \begin{tabular}[c]{@{}c@{}}$7\times10^{16}$\\ (120 nmol)\end{tabular} & noninductive \\
\hline

Microcoil & \cite{S_mcDowellMicrocoil}   & $^1$H in water & 1 T & yes & 81 nL   & \begin{tabular}[c]{@{}c@{}}$3\times10^{16}$\\ (50 nmol)\end{tabular}  & inductive    \\
\hline

Microslot & \cite{S_suterMicroslot}   & \begin{tabular}[c]{@{}c@{}}$^1$H in 215 mM of\\ sucrose in D$_2$O\end{tabular} & 11.7 T & yes & 10.6 nL & \begin{tabular}[c]{@{}c@{}}$3\times10^{14}$\\ (0.5 nmol)\end{tabular} & inductive    \\
\hline

GMR & \cite{S_gmr_nmr}   & $^1$H in water & 0.3 T & yes & 62 pL   & \begin{tabular}[c]{@{}c@{}}$1\times10^{17}$\\ (170 nmol)\end{tabular} & noninductive \\
\hline

AMR & \cite{S_amr_nmr}   & $^1$H in water & \begin{tabular}[c]{@{}c@{}}17 T prepolarize\\ 0 T detection\end{tabular}  & yes & 1 pL    & \begin{tabular}[c]{@{}c@{}}$3\times10^{14}$\\ (0.5 nmol)\end{tabular} & noninductive \\
\hline

Single NV & \cite{S_rugar2015}   & $^1$H in PPMA & 40 mT & no & $<$1 fL   & N/A & noninductive \\

\hline

\end{tabular}
\caption{\label{fig1bTable} Overview of NMR techniques for small volumes. Here we list the number of spins detectable with SNR = 3 after 1 s of averaging. %**** what about DeVience ensemble and Lovchinsky repetitive readout?
}
\end{table}
Table \ref{fig1bTable} summarizes the NMR techniques indicated in Fig.~1a in the main text. Compared to NV NMR spectroscopy, the Microcoil/microslot/cryo-probe techniques have finer frequency resolution and can distinguish $^1$H at different nuclear sites within a molecule. The determination of $\rho_{min}$ under these techniques often focuses on a single site within a molecule, such as the anomeric proton in sucrose, so that the analyte concentration is the same as the detected spin concentration. For current implementations of NV NMR, chemical shifts are indistinguishable, so all nuclei of a given species contribute equally to the signal. Thus, when determining $\rho_{min}$ in our technique, we must compute the concentration of all spins of the target species within the analyte, not just the concentration of analyte molecules.
\begin{figure}[ht]
\begin{center}
\includegraphics[width=.4\textwidth]{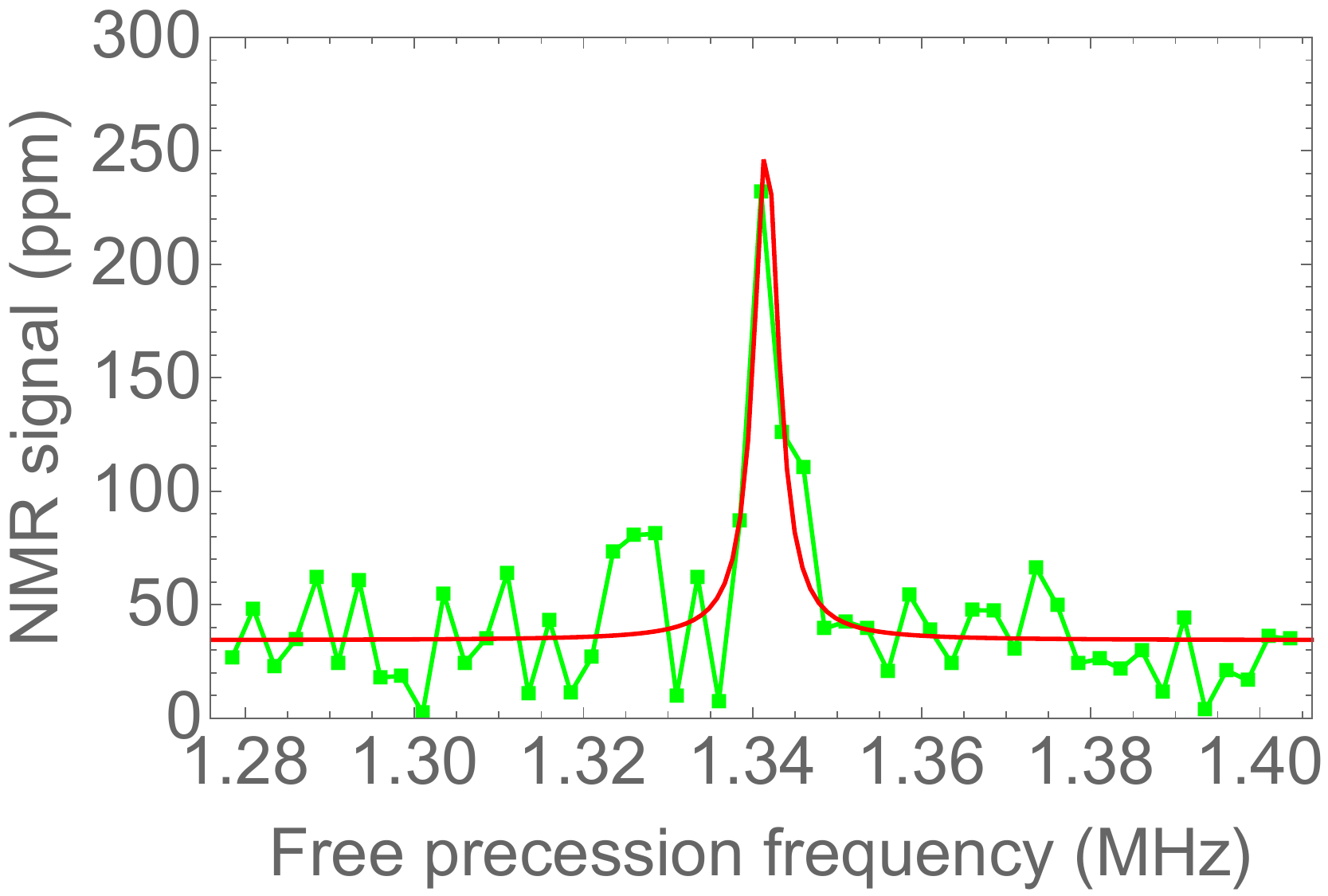}
\end{center}
\caption{\label{narrowCorrSpec}
A $^1$H NMR spectrum for a correlation spectroscopy measurement with nanograting chip UNM6 (not listed in the above tables; 33\%:67\% diamond/gap duty cycle 40 keV implant, $5\times10^{13}$ $^{15}$N$^+$/cm$^2$, $\pm 5^{\circ}$ degree implant angle) with microscope immersion oil ($B_0 = 31.5~$mT). This is the narrowest NMR line we measured in our experiments, with a full-width-at-half-maximum of 3.5 kHz.
}
\end{figure}
\section{Narrowest-observed spectral resolution}
The NV $T_1$ time sets the best-case correlation spectroscopy linewidth to $(\pi T_1)^{-1} \approx 100$ Hz FWHM, which is often unachievable due to molecular diffusion broadening. Figure \ref{narrowCorrSpec} shows the narrowest NMR peak, 3.5 kHz FWHM, we observed using a 40 keV implant nanograting chip (not included in the main analysis presented).  This especially narrow linewidth may be due to the ``hardened oil effect" causing reduced molecular diffusion near the diamond surface \cite{S_merilesDiffusion}, and emphasizes that restricting diffusion is important to fully exploit correlation spectroscopy with shallow NVs.

%\bibliography{gratingBiblioSI}
%

\end{document}